\documentclass[pdflatex,sn-mathphys-num]{sn-jnl}% Math and Physical Sciences Numbered Reference Style
\usepackage{graphicx}%
\usepackage{multirow}%
\usepackage{amsmath,amssymb,amsfonts}%
\usepackage{amsthm}%
\usepackage{mathrsfs}%
\usepackage[title]{appendix}%
\usepackage{xcolor}%
\usepackage{textcomp}%
\usepackage{manyfoot}%
\usepackage{booktabs}%
\usepackage{algorithm}%
\usepackage{algorithmicx}%
\usepackage{algpseudocode}%
\usepackage{listings}%
\usepackage{tikz-cd}
\usepackage{adjustbox} %for command \adjustbox, which I used to scale tikz-cd diagrams.
\usepackage{subcaption} %for subfigures
\usepackage{xfrac} %for fraction written as 1/2

\usepackage{censor} % hiding for peer review
\StopCensoring
\newcommand{\R}{\mathbb{R}}
\newcommand{\N}{\mathbb{N}}

\newcommand{\X}{\mathcal{X}}
\newcommand{\Y}{\mathcal{Y}}
\newcommand{\U}{\mathcal{U}}
\newcommand{\Z}{\mathcal{Z}}

\renewcommand{\a}{\bar{\alpha}}
\renewcommand{\b}{\bar{\beta}}
\newcommand{\h}[2]{h^{#1,\mspace{1mu}#2}}

\newcommand{\Viterbi}[1]{\the\numexpr#1-0\relax-Viterbi}

\newcommand{\Max}[1]{\raisebox{0.5ex}{\scalebox{0.8}{$\displaystyle \max_{#1}\;$}}}

\DeclareMathOperator*{\argmax}{arg\,max}
\DeclareMathOperator*{\argmin}{arg\,min}

\theoremstyle{thmstyleone}%
\theoremstyle{thmstyletwo}%

\theoremstyle{thmstylethree}%

\begin{document}

\title[Article Title]{Branch-and-bound method for calculating Viterbi path in triplet Markov models}


%Hidden for peer-review
\author*[1]{\fnm{Oskar} \sur{Soop}}\email{oskar.soop@ut.ee}

\author[1]{\fnm{J\"uri} \sur{Lember}}\email{jyril@ut.ee}

\affil*[1]{\orgdiv{Institute of Mathematics and Statistics}, \orgname{University of Tartu}, \orgaddress{\street{Narva mnt 18}, \city{Tartu linn}, \postcode{51009}, \state{Tartumaa}, \country{Estonia}}}


\abstract{We consider a bivariate, possibly non-homogeneous, finite-state Markov chain $(X,U)=\{(X_t,U_t)\}_{t=1}^n$. We are interested in the marginal process $X$, which  typically is not a Markov chain. The goal is to find a realization (path) $x=(x_1,\ldots,x_n)$ with maximal probability $P(X=x)$. If $X$ is Markov chain, then such path can be efficiently found using the celebrated Viterbi algorithm. However, when $X$ is not Markovian, identifying the most probable path---hereafter referred to as the \textit{Viterbi path}---becomes computationally expensive. In this paper, we explore the branch-and-bound method for finding Viterbi paths. The method is based on the lower and upper bounds on maximum probability $\max_x P(X=x)$, and the objective of the paper is to exploit the joint Markov property of $(X,Y)$ to calculate possibly good bounds in possibly cheap way.\\
This research is motivated by decoding or segmentation problem in triplet Markov models. A triplet Markov model is trivariate homogeneous Markov process $(X,U,Y)$. In decoding, a realization of one marginal process $Y$ is observed (representing the data), while $X$ and $U$ are latent processes. The process $U$ serves as a nuisance variable, whereas $X$ is the process of primary interest. Decoding refers to estimating the hidden sequence $X$ based solely on the observation $Y$. Conditional on $Y$, the latent processes $(X, U)$ form a non-homogeneous Markov chain. In this context, the Viterbi path corresponds to the maximum a posteriori (MAP) estimate of $X$, making it a natural choice for signal reconstruction.}

\keywords{
%Triplet Markov models, 
Pairwise Markov models,
Hidden Markov models,
%Viterbi path, 
Maximum a posteriori estimation, 
maximal marginal,
Branch and bound 
%Power sum bounds, 
%Samuelson-type inequalities,
%Non-Markovian inference, 
%Sequence segmentation, 
%NP-hard optimization, 
%Markov chain marginalization
}



\maketitle

\section{Introduction}\label{sec1}
In this article, we consider the computational methods for finding the Viterbi, or equivalently, the maximum likelihood path, for a
non-homogeneous finite-state Markov chain $(X,U)=\{(X_t,U_t)\}_{t=1}^n$. In particular, we are interested in the marginal process $X$, which generally does not inherit the Markov property. The goal is to find a realization (path) $x=(x_1,\ldots,x_n)$ that maximizes the probability $P(X=x)$.

If $X$ forms a Markov chain, such a path can be efficiently found using the celebrated Viterbi algorithm. However, when $X$ is not Markovian, identifying the most probable path---hereafter referred to as the \textit{Viterbi path}---becomes computationally expensive. We refer to this task as the \textit{maximal marginal problem}.

The motivation for the maximal marginal problem stems from the segmentation task in triplet Markov models (TMMs). A TMM is a trivariate homogeneous Markov process $(X,U,Y)$. In segmentation, one observes a realization of a single marginal process $Y$ (representing the data), while $X$ and $U$ are latent processes. Typically, $U$ serves as a nuisance variable, whereas $X$ is the process of primary interest. A key property of TMMs is that, conditioned on a realization of $Y$, the joint process $(X, U)$ becomes a non-homogeneous Markov chain. Finding the maximum a posteriori path -- a standard solution of the segmentation problem -- is then exactly the maximal marginal problem.

 The maximal marginal problem (i.e., finding a Viterbi path without Viterbi algorithm) is known to be a NP-hard problem. In this paper, we solve that problem via branch-and-bound method. This method relies on upper and lower bounds for $\max_{x}P(X=x)$, and its overall computational complexity depends on the tightness of these bounds and the efficiency with which the bounds can be computed.
 
 We consider several types of bounds. Besides the trivial bounds, we consider the power sum bounds, Samuelson type bounds, swapped max-sum bounds and $m$-Viterbi approximations. All these bounds will be described in detail in Section \ref {sec:exact_algorithms} and also analyzed in Appendix.
 Roughly speaking, the power-sum upper bound is based on inequality  $\max_{x}P(X=x)\leq \big(\sum_x P(X=x)^r\big)^{\frac{1}{r}}$. Here $r\in \mathbb{N}$ is the power, the bigger $r$, the sharper and more complex is the inequality. Samuelson-type bounds combine the power-sums for $r=1,2$. Swapped max-sum bounds exploits the fact the changing the order of summation and maximization increases the objective. In  $m$-Viterbi approximation, the process $X$ is approximated by a $m$-th order Markov chain, and the maximum-probability path of that approximation, let that be $\hat{x}$, is found. The path $\hat{x}$, together with its probability $P(X=\hat{x})$, can then be found by modified Viterbi algorithm. So we have a lower bound $P(X=\hat{x})\leq \max_{x}P(X=x)$. In order to calculate all these above-mentioned bounds efficiently, the joint Markov property of $(U,X)$ is used.

 The paper is organized as follows. In Subsection~\ref{sec:MMM}, we introduce the stochastic models under consideration---pairwise Markov models (PMMs) and triplet Markov model---and review their key properties and examples. Subsection~\ref{subsection2} gives a very short overview of segmentation problem in statistical learning framework, providing motivation for the maximal marginal problem. In subsection~\ref{sec:dynamic_programming}, we recall the basic dynamic programming algorithms such as the Viterbi and forward-backward recursion, which form the foundation for our methods.
 In Section~\ref{sec:exact_algorithms}, we describe the branch-and-bound algorithm together with a variety of bounds. Section~\ref{sec:experiments} presents the empirical results comparing the bounds. The simulations assure that branch-and-bound approach is clearly more efficient than exhaustive search. However, they do not decisively indicate superiority of any particular upper bound. Among the tested bounds, the $m$-Viterbi approximation generally provides the best lower bound.
 
 In Appendices we provide some proofs, algorithms and additional information about the considered bounds. In particular, some formulas for computation are derived. Given the strong empirical performance of the $m$-Viterbi approximation, we examine it in greater depth. In particular, we present counterexamples challenging common intuitions: (1) that increasing $m$ always improves the approximation, and (2) that a Markov approximation always yields a path with positive probability. These findings align with our empirical results, underscoring that the effectiveness of the $m$-Viterbi approximation does not necessarily improve with larger $m$. These counterexamples align with our empirical results, underscoring that show that the goodness of $m$-Viterbi approximation need not necessarily increase with $m$.

\section{Preliminaries}

%-----------------------------------------------
\subsection{The multiple Markov  models}\label{sec:MMM}
%-----------
A {\it \textbf{pairwise Markov model} (\textbf{PMM})} is a
bivariate Markov chain $\{Z_t\}_{t \geq
1}=\{(X_t,Y_t)\}_{t \geq 1}$  taking values on  $\Z\subseteq \X
\times \Y$, where $\X=\{1,\ldots,|\X|\}$  is a finite set, typically referred to as the {\it state-space} and $\Y$ is a possibly uncountable set.
Process
$Y=\{Y_t\}_{t \geq 1}$ is seen as the observed sequence and
$X=\{X_t\}_{t \geq 1}$ is typically seen as the hidden or latent variable
sequence, often referred to as  {\it the signal process}.
Generally, neither $Y$ nor
$X$ is a Markov chain, although for special cases they might be. In
many practical models  the signal process $X$ remains to be a Markov
chain. However, for every PMM, conditionally on the realization of
$X$ (resp. $Y$), the $Y$  (resp. $X$) is always an
non-homogeneous Markov chain (see the last paragraph in Subsection \ref{subsection2}).
When $\Y$ is countable, then $Z$ has countable state space $\Z$, hence  it is specified by its transition matrix and the distribution on $Z_1$. For examples of such PMM's see \cite{hyb}. When $\Y$ is uncountable, then $Z$ is specified by a transition kernel which is assumed to have a density with respect to the product measure $\mu\times c$, where $\mu$ is a reference measure on $\Y$ (typically Lebesgue measure when $\Y=\mathbb{R}^d$) and $c$ is the counting measure on $\Y$. We also assume that $Z_1$ has a density with respect to $\mu\times c$ and then for every $n$ the vector $(Z_1,\ldots,Z_n)$ has density with respect to $\mu\times c$ as well, see \cite{sova1,sova2,sova3} for details.
\\\\
PMMs are a very large and flexible class of models including many important subclasses. Probably the most-known {non-trivial} PMM is a {\it hidden Markov model} ({HMM}). The characteristic features of a HMM is that the signal process is Markov and, conditionally on the signal, the observations are independent. This particular property is very restrictive in many applications. So, PMM allows to have observations (conditionally) dependent and the signal process not Markov so that the joint process remains to be Markov. This property -- being jointly Markov -- solely defines a PMM and makes the classical HMM-tools like forward-backward and Viterbi algorithms also possible for PMMs (see Section \ref{sec:dynamic_programming}). Formally, of course, one can consider every PMM $(X,Y)$ as a HMM  $((X,Y),Y)$, i.e. the signal process in $(X,Y)$ and observations are just projections (and emission distributions are degenerate). Therefore this isomorfism between HMMs and PMMs is purely theoretical and as much as applications are concerned, the PMM is a way larger class of models in comparison with HMM. For examples, classification, properties, applications and theoretical results of various PMM-models, see
 \cite{P03, P04, P04b, Pzebra,P12,P13, P16, gorynin18, sova1,sova2,sova3,hyb}.

A {\it \textbf{triplet Markov model} (\textbf{TMM})} is trivariate Markov chain  $\{Z_t\}_{t \geq
1}=\{(X_t,Y_t,U_t)\}_{t \geq 1}$  taking values on  $\Z\subseteq \X
\times \Y \times {\cal U}$, where, $\X$ and $\Y$ are as previously, and ${\cal U}=\{1,\ldots,|\U|\}$ is a finite set. Rather than the dimension of the state space, in classifying a Markov chain as PMM or TMM, the roles of $X,Y,U$ are important. In TMM, typically, $Y$ stands for the observed sequence, the $X$-process is the signal of interest and the additional $U$-process stands for an auxiliary or nuisance process that is neither observed nor of interest, but necessary for modeling. Again, formally every TMM is a PMM when considering two marginal processes as one. For example a TMM $(X,Y,U)$ is a PMM $(Y,V)$, where $V=(X,U)$.  So, when
when $\Y$ is countable, then $Z$ has countable state space $\Z$ and  is specified by its transition matrix and the distribution on $Z_1$, $\Y$ is uncountable, then $Z$ is specified by a transition kernel just like in PMM case.
%-------------------------
In general the (one or two dimensional) marginal processes of a TMM are not Markov ones, but sometimes, depending on the model, it might be so. In the present 
paper, we consider the case when the two-dimensional marginal $(X,Y)$ is not necessarily a Markov chain, because otherwise the objective of this article -- Viterbi path -- could simply be obtained by a Viterbi algorithm.
\\\\
%-------
Probably the most commonly used TMM is a PMM with independent noise, where $(V,Y)$ is a HMM, with $V=(X,U)$ being a PMM and $Y$ is the observation process, the distribution of  $Y_t$ depending solely on $X_t$;   see \cite{P04b,P07,Pzebra,BCA06,Deep23} for  examples in image segmentation, \cite{Biv13,Biv16} in spectrum sensing  and \cite{P19} in activity monitoring. Since every discrete semi-Markov model $X$ can be modelled as a marginal of a PMM, see e.g. \cite{hyb}, also the hidden semi-Markov model $(X,Y)$ can be modeled as a marginal process of  a  TMM, \cite{Psemi,Psemi10,P23}. In all these models, actually the pair $(X,Y)$ is of interest, but since $X$ is not a Markov chain, then $(X,Y)$ is not a HMM (and not a PMM), so all HMM-tools are useless. With the nuisance process $U$, however, the triplet $(X,U,Y)$ is a Markov model, and the tools might apply.
\\\\
Finally, let us remark that every Markov model can be considered a specific instance of Bayesian networks (or more generally, of probabilistic graphical models; see book \cite{Book:ProbabilisticGraphicalModels} for an introduction). Conversely, every Bayesian network can be modeled as a Markov model, with caveat that the state space is generally non-constant in time. This "Markovification" can be done with a modified version of topological sort as seen in Figure \ref{fig:Markovification}. There are multiple ways to transform the Bayes network into a Markov chain and this selection can influence the algorithms' performance.
As an illustrative example why this selection matters, assume that processing an arrow between $X_t$ and $X_{t+1}$ in Markov chain would take $|\X_t||\X_{t+1}|$ steps (e.g., due to evaluating a transition matrix of that size).  In case of the chain in Figure \ref{subfig:Markov} it would take $|\mathcal{A}\times\mathcal{B}\times\mathcal{C}| + |\mathcal{C}\times\mathcal{D}\times\mathcal{E}|$ steps to process all arrows. Alternatively consider Markov chain, where node $A$ is grouped with $B$, node $D$ is grouped with $C$ and $E$ is alone. In that case it would take $|\mathcal{A}\times\mathcal{B}\times\mathcal{C}\times\mathcal{D}| + |\mathcal{C}\times\mathcal{D}\times\mathcal{E}|$ steps to process all arrows.

\usetikzlibrary{positioning, shapes, fit, arrows.meta}

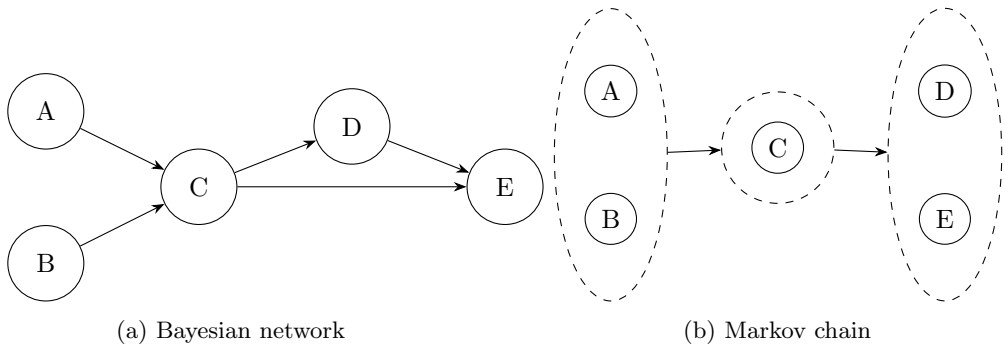
\begin{figure}[ht]
\centering

% First diagram: original DAG
\begin{subfigure}[t]{0.45\textwidth}
\centering
\begin{tikzpicture}[
  ->, >=Stealth,
  node distance=1cm and 1cm,
  every node/.style={draw, circle, minimum size=1cm, inner sep=0pt}
]
\node (A) {A};
\node (B) [below=of A] {B};
\node (C) [right=of A, yshift=-1cm] {C};
\node (D) [right=of C, yshift=0.8cm] {D};
\node (E) [right=of D, yshift=-0.8cm] {E};

\draw (A) -- (C);
\draw (B) -- (C);
\draw (C) -- (D);
\draw (C) -- (E);
\draw (D) -- (E);
\end{tikzpicture}
\caption{Bayesian network}
\label{subfig:Bayes}
\end{subfigure}
\hfill
% Second diagram: Markovified
\begin{subfigure}[t]{0.45\textwidth}
\centering
\begin{tikzpicture}[
  node distance=1cm and 1.5cm,
  every node/.style={draw},
  ell/.style={draw, ellipse, inner sep=5pt, dashed, fit=#1},
  ->, >=Stealth
]

\node[circle] (A) {A};
\node[circle, below=of A] (B) {B};
\node[circle, right=of A, yshift=-0.75cm] (C) {C};
\node[circle, right=of C, yshift=0.75cm] (D) {D};
\node[circle, below=of D] (E) {E};

\node[ell={(A)(B)}] (ABgroup) {};
\node[ell={(C)}] (Cgroup) {};
\node[ell={(D)(E)}] (DEgroup) {};

\draw (ABgroup) -- (Cgroup);
\draw (Cgroup) -- (DEgroup);
\end{tikzpicture}
\caption{Markov chain}
\label{subfig:Markov}
\end{subfigure}

\caption{"Markovification" of the Bayes network}
\label{fig:Markovification}
\end{figure}

The distinction in designing inference algorithms for Bayesian networks and TMMs, while they are technically the same, is that usually Bayesian networks are "wide" and TMMs are "long". Formally, this means that the Bayesian networks can have arbitrarily large tree-width, while TMMs have bounded tree-width (at most 3), but potentially long time horizons. Our focus is on non-homogeneous Markov chains, where state space is constant in time. This means that there is no need for advanced selection strategies for the order of variables, such as is the case with the Bayes networks -- we can simply start processing Markov chains from the start or from the end.

\subsection{The segmentation problem}\label{subsection2}
In the present paper, we  consider a TMM $(U,X,Y)$ such that the marginal pair $(X,Y)$ is not a PMM. In what follows, we consider the finite time-horizon $n$, so that $X=(X_1,...,X_n)$, $Y=(Y_1,...,Y_n)$ and $U=(U_1,...,U_n)$. Throughout the article we shall denote by lower indices in $a_{1:n}$ the vector $(a_1,a_2,\ldots, a_n)$ and $a_{s:t}$ ($1\leq s<t\leq n$) stands for the segment $(a_s,a_{s+1},\ldots,a_t)$ of $a_{1:n}$. These lower indices will usually refer to \textit{time} and when we want to denote vectors of state space we use upper indices as in $a^{1:r}$, which refers to vector $(a^1, a^2, \dots, a^r)$. The random variables $X_t$, $Y_t$ and $U_t$ will take the values in the sets $\X$, $\Y$ and $\U$, respectively, for every $t$.\\\\
%----------------
With a slight abuse of notation, in what follows the letter $p$ will be used to denote various joint and conditional densities and probabilities, for example $p(x_{1:n},y_{1:n})$, where $x_{1:n}\in \X^n$ and $y_{1:n}\in \Y^n$ stands for density of $(X_{1:n},Y_{1:n})$ and $p(x_{1:n}|y_{1:n})$,  
$p(x_t,y_t|x_{t-1},y_{t-1})$ stand for conditional densities and so on. We shall  denote $P(X_{u:v}=s_{u:v}|Y_{1:n}=y_{1:n})$ as
$p(x_{u:v}=s_{u:v}|y_{1:n})$ or $p(s_{u:v}|y_{1:n})$ and $P(X_{u:v}=s_{u:v}|X_{u-1}=i,Y_{1:n}=y_{1:n})$ as $p(x_{u:v}=s_{u:v}|x_{u-1}=i,y_{1:n})$. 
%----
%In what follows,  we assume that a realization $y_{1:n}$ of $Y_{1:n}$ -- the %observations -- are known,  but the  realizations of $X$ and $U$ are not observed.
%---------
\paragraph{Elements of risk-based segmentation theory.} Let $(X,Y)=(X_{1:n},Y_{1:n})$ be any bivariate process, not necessarily PMM or marginal of a TMM. We assume a realization $y_{1:n}$ of $Y$ is known.
The {\it segmentation (denoising, decoding) problem} consists of estimating/predicting  the
unobserved realization of the underlying process $X_{1:n}$ given
observations $y_{1:n}$. Formally, we are looking for a mapping
$g:{\cal Y}^n \to {\cal X}^n$ called a {\it classifier} or {\it
decoder}, that maps every sequence of observations into a state
sequence. The best classifier $g$ is often defined via a {\it loss function} $L: {\cal X}^n\times {\cal X}^n \to [0,\infty],$ where
$L(x_{1:n},s_{1:n})$ measures the loss when the actual state sequence is $x_{1:n}$ and the estimated sequence is $s_{1:n}$. For
any state sequence $s_{1:n}\in {\cal Y}^n$, the expected loss for given the observations $y_{1:n}$  is called {\it conditional risk}:
%------------------------
\[\label{risk}
R(s_{1:n}|y_{1:n}):=\sum_{x_{1:n}\in
{\cal X}^n}L(x_{1:n},s_{1:n})p(x_{1:n}|y_{1:n}).\]
The best classifier maps any $y_{1:n}$ to a state sequence minimizing the conditional risk:
$$g^*(y_{1:n})=\argmin_{s_{1:n}\in {\cal X}^n}R(s_{1:n}|y_{1:n}).$$
For an overview of risk-based segmentation with HMMs and PMM's see
\cite{seg,intech,chris,hyb}. The two most common loss functions used in practice are the global loss function $L_{\infty}$,
\[\label{symm-n}
L_{\infty}(x_{1:n},s_{1:n}):=\left\{
               \begin{array}{ll}
                 1, & \hbox{if $x_{1:n}\ne s_{1:n}$,} \\
                 0, & \hbox{if $x_{1:n}=s_{1:n}$,}
               \end{array}
             \right.
\]
and the local loss function
\begin{equation}\label{kadu-p}
  L_1(x_{1:n},s_{1:n}):=\sum_{t=1}^n I(x_t\ne s_t).
\end{equation}
The conditional risk corresponding to $L_{\infty}$  is
$1-p(x_{1:n}=s_{1:n}|y_{1:n})$, thus the best classifier finds the path with  the maximum posterior probability:
\begin{equation}\label{map}
x^*_{1:n}:=\argmax_{x_{1:n}\in \X^n}p(x_{1:n}|y_{1:n}).\end{equation}
%------ \quad \text{-- Viterbi alignment or Viterbi path}
Any state path (\ref{map}) (it is not necessarily unique) is called the  {\it maximum a posteriori (MAP) path.} When $(X,Y)$ is a PMM, then MAP path is also referred to as {\it Viterbi path} or {\it Viterbi alignment}, due to the Viterbi algorithm that is used to find it. In what follows, we shall refer to any MAP path as a Viterbi path even when it cannot be found via Viterbi algorithm.
\\\\
The conditional risk corresponding to  $L_1$ in (\ref{kadu-p}) is the expected number of misclassification errors and  can be calculated as follows:
$$n-\sum_{t=1}^np(x_t=s_t|y_{1:n}).$$
Hence, the best classifier corresponding to $L_1$ finds the path with minimal expected number of misclassification errors as follows: 
\begin{equation}\label{pmap}
x^*_t=\argmax_{x_t\in {\cal X}}p(x_t|y_{1:n}),\quad t=1,\ldots,n.
\end{equation}
We will call any such $x^*_{1:n}$ a {\it pointwise maximum a posteriori (PMAP)} path. In PMM literature often the name {\it maximum posterior mode (MPM)} is used; see, for example, \cite{P03,P04,Pzebra,P12,gorynin18}. Unlike the Viterbi path, the PMAP path can be computed independently at each time step $t$. Since the state space $\X$ is typically small in practical applications, the maximization in \eqref{pmap} is computationally trivial. Therefore, computing the PMAP path reduces to evaluating the \textit{smoothing probabilities} $p(x_t \mid y_{1:n})$, which can be efficiently obtained using the classical forward-backward algorithm (see Section~\ref{sec:dynamic_programming}).
\\\\
Although commonly used and by far most popular, the standard classifiers suffer from notable shortcomings. The Viterbi path often lacks accuracy and exhibits systematic errors \cite{sova1,sova2}. On the other hand, the PMAP path might have very low or even zero probability (inadmissible). To overcome those deficiencies, in \cite{seg}, a class of {\it hybrid paths} were introduced. In the standard form any  hybrid path is a solution of the following (combined) problem
$$\max_{x_{1:n}\in {\cal X}^n}\bigg[\sum_{t=1}^n \ln p(x_t|y_{1:n})+C\ln p(x_{1:n}|y_{1:n})\bigg],$$
where $C\geq 0$ is a regularization constant. By varying $C$, one can interpolate between these two extremes, often achieving a balance that combines the desirable properties of both paths --- namely, high marginal accuracy and high joint probability; see, e.g., \cite{hobolth}. For PMMs, hybrid paths can be computed using a Viterbi-like dynamic programming algorithm. However, for TMMs, finding hybrid paths is computationally as challenging as computing the Viterbi path. 
%-----
\paragraph{Segmentation with TMMs and Statement of the Maximal Marginal Problem.}
Let $(X, Y)$ be a PMM. From a segmentation perspective, the key property of a PMM is that, conditioned on the observations $Y_{1:n} = y_{1:n}$, the hidden process $X_{1:n}$ is a non-homogeneous Markov chain:
\[
p(x_{t+1} \mid x_{1:t}, y_{1:n}) = p(x_{t+1} \mid x_t, y_{t:n}).
\]
This well-known and easily proven property (see, e.g., \cite{P03,hyb}) underpins all dynamic programming algorithms used in PMMs. It ensures that various decoding strategies, including hybrid paths, the Viterbi path, and the PMAP path, can be efficiently computed.
\\\\
%-------------------
Now, let $(X, U, Y) = (X_{1:n}, U_{1:n}, Y_{1:n})$ be a TMM. Our interest lies in the marginal pair $(X, Y)$, which no longer forms a PMM. This raises the question: how can standard (e.g., PMAP and Viterbi) and hybrid classifiers be computed in this setting?
First, observe that the joint process $(V, Y)$, where $V = (X, U)$, constitutes a PMM. Therefore, conditioned on the observations $Y_{1:n} = y_{1:n}$, the process $(X, U)$ is a non-homogeneous PMM. As a result, the joint smoothing probabilities $p(x_t, u_t \mid y_{1:n})$ can be efficiently computed. This allows for computing the marginal smoothing probabilities for $(X, Y)$ via
\[
p(x_t \mid y_{1:n}) = \sum_{u_t \in {\cal U}} p(x_t, u_t \mid y_{1:n}),
\]
and thus the PMAP path (as defined in equation~\eqref{pmap}) can be readily obtained.
In practice, and throughout the literature on TMM-based segmentation, PMAP classifiers are preferred due to their computational tractability. However, computing the Viterbi path for the marginal model $(X, Y)$ is substantially more challenging and is the main goal of the present work.
Since the observations $y_{1:n}$ are fixed, and the conditional process $(X, U)| y_{1:n}$ is a non-homogeneous PMM, we simplify notation by omitting \(y_{1:n}\) in what follows. The central problem considered in this article—referred to as the \textbf{maximal marginal problem}—is formulated in the following general form: given a non-homogeneous PMM \((X, U)\) with finite state spaces, find the Viterbi path \(x_{1:n}\) that maximizes the marginal probability
\begin{equation}\label{problem}
p(x_{1:n}) = \sum_{u_{1:n} \in {\cal U}^n} p(x_{1:n}, u_{1:n}).
\end{equation}
The solution to the maximal marginal problem is not necessarily unique; the branch-and-bound method introduced in Section~\ref{sec:branch-and-bound} identifies all the solutions.
\\\\
In the literature on HMMs or the more general framework of probabilistic graphical models, the maximal marginal problem---or its slight variation, the Viterbi path problem---is also called {\it most likely string} \cite{goodman1998parsinginsideout}, {\it consensus string} \cite{consensus_string}, {\it max-sum-product} \cite{ApproximatingMaxSumProduct} and {\it maximum a posteriori} (MAP) \cite{Solving_MAP_Exactly_using_Systematic_Search} problem. For Bayesian networks, this problem is well known to be NP-hard to solve exactly \cite{MAPisNPHardInBN}, and even to approximate \cite{MAPisNPHardApproxInBN}. A less widely known fact is that it is also NP-hard to solve in the case of HMMs \cite{goodman1998parsinginsideout,consensus_string}, and therefore also for PMMs. In the article \cite{consensus_string}, it is shown that the well-known NP-hard problem called the maximum clique problem reduces to the maximal marginal problem in HMMs. I.e. it is possible to compute the size of the maximum clique in an undirected graph by computing maximal marginal probability by constructing a specific HMM in polynomial time.
Furthermore, using the inapproximability results of the maximum clique problem \cite{max-clique} and the aforementioned reduction, it can be shown that there is no polynomial time algorithm that approximately solves the maximal marginal problem within a factor of $O(n^{\sfrac{1}{2}-\varepsilon})$ for any $\varepsilon>0$ unless P = NP.
Although the problem is NP-hard, it is possible to significantly improve upon the exhaustive search by using a branch-and-bound algorithm. In this article, we present a branch-and-bound algorithm alongside a  variety of bounds for solving the maximal marginal problem. We also present some heuristic and approximation methods. For simplicity, in the present article we deal with Viterbi paths only. However, the obtained methods apply also for finding the hybrid paths. 
\\\\
The model considered in the present article is connected to the Bayesian PMMs approach, since the nuisance  process $U$ can be regarded as a random parameter. Indeed, given $U$, the conditional process $(X,Y)|U=u_{1:n}$ is a non-homogeneous PMM. However, when summing $u_{1:n}$ out, the Markov property no longer holds. This is analogous to the Bayesian case: given a parameter $\theta$, the process $(X,Y)|\theta$ is a PMM, but once the parameter is integrated out, the Markov property is lost \cite{Metron}.
Nevertheless, the resemblance ends there. In the typical (parametric) Bayesian setting—where the parameter space is uncountable—there are generally no computationally efficient methods for calculating the PMAP path. Iterative algorithms for approximating the Viterbi path in Bayesian HMMs have been proposed, for instance, in \cite{Metron,case}.
%--------------------------------
\subsection{Dynamic programming tools}\label{sec:dynamic_programming}
\paragraph{Viterbi algorithm.} Recall our problem: given (non-homogeneous) PMM $(X,Y)$ find the (any, if many) path that maximizes the probability $p(x_{1:n})=P(X=x_{1:n})$, where $p(x_{1:n})$ is as in (\ref{problem}). If the process $X$ is a Markov chain then maximization can be solved by dynamic programming
\begin{align*}
\max_{x_{1:n}} p(x_{1:n}) &= \max_{x_1} \max_{x_2} \ldots \max_{x_n} p(x_1)p(x_2|x_1)\ldots p(x_n|x_{n-1}) \\&=
\max_{x_1} p(x_1) \max_{x_2} p(x_2|x_1) \ldots \max_{x_n} p(x_n|x_{n-1}).
\end{align*}
That observation is the basis of  the celebrated {\it Viterbi algorithm}: with
$$\delta_t(x_{t-1}):=\max_{x_{t+1:n}}p(x_{t:n}|x_{t-1}),\quad t=2,\ldots,n-1,$$ the (backward) Viterbi recursion is
\begin{align}\label{VAb}\delta_{t}(x_{t-1})=\max_{x_{t}}p(x_{t}|x_{t-1})\delta_{t+1}(x_t).\end{align}
The maximum value is $\max_{1:n}p(x_{1:n})=\max_{x_1}p(x_1)\delta_2(x_1)$ and the MAP path can be found by recording the argmax-values in every step of iteration and backtracking from beginning. Observe that Viterbi algorithm could also be applied in reverse direction, where 
\begin{equation*}\delta_1(x_1):=p(x_1),\quad \delta_t(x_t):=\max_{x_{1:t-1}}p(x_{1,t-1},x_t),\quad t=1,\ldots,n.\end{equation*}
The (forward) Viterbi recursion is then
\begin{equation}\label{VAf}
\delta_{t+1}(x_{t+1})=\max_{x_t}p(x_{t+1}|x_i)\delta_t(x_t)\end{equation}
and $\max_{1:n}p(x_{1:n})=\max_{x_n}\delta_n(x_n).$
In order to apply the methods described in the paper, both versions of Viterbi algorithm are used. Observe that they could be applied simultaneously -- use forward Viterbi algorithm to compute $\max_{x_{1:t-1}}p(x_{1,t-1},x_t)$ and backward Viterbi algorithm to compute $\max_{x_{t+1:n}}p(x_{t+1:n}|x_t)$, where $t$ is a fixed time. Multiplying these probabilities and maximizing over all possible values of $x_t$ gives the $\max_{x_{1:n}} p(x_{1:n})$; backtracking from both directions gives a Viterbi path.
The Viterbi algorithm reduces the number of operations from $O(|{\cal X}|^n)$ to 
$O(n\vert\X\vert^2)$.
\\\\
Unfortunately, in our case $X$ is not a Markov process, so Viterbi algorithm  cannot be applied. It could applied to the joint process  $(X,U)$, resulting the maximum probability pair
\begin{equation}\label{ux}
(\hat{x}_{1:n},\hat{y}_{1:n})=\argmax_{(x_{1,n},u_{1:n})}p(x_{1,n},u_{1:n}).
\end{equation}
The marginal $\hat{x}_{1:n}$ is typically not a Viterbi path, it will be used as an estimate.

\paragraph{Forward-backward algorithms.} Since $(X,U)$ is a Markov chain then marginalization (i.e. finding $p(x_{1:n})$) can be solved by analogous  trick
\begin{align*}
\sum_{u_{1:n}} p(x_{1:n},u_{1:n}) &= \sum_{u_1} \sum_{u_2} \ldots \sum_{u_n} p(x_1,u_1)p(x_2,u_2|x_1,u_1)\ldots p(x_n,u_n|x_{n-1},u_{n-1}) \\&=
\sum_{u_1} p(x_1,u_1) \sum_{u_2} p(x_2,u_2|x_1,u_1) \ldots \sum_{u_n} p(x_n,u_n|x_{n-1},u_{n-1}).
\end{align*}
This is the basis of belief propagation algorithm, which in Markov models terminology is typically referred to as the backward-algorithm. It goes as follows: given the path $x_{1:n}$, define 
$$\beta_t(u_t):=p(x_{t+1:n}|x_t,u_t),\quad \beta_n(u_n)\equiv 1.$$
The backward recursion is 
\begin{equation}\label{beta}
\beta_{t-1}(u_{t-1})=\sum_{u_t} p(x_t,u_t|x_{t-1},u_{t-1})\beta_t(u_t).\end{equation}
Hence, the probability of the path is $p(x_{1:n})=\sum_{u_1}p(x_1,u_1)\beta_1(u_1)$. Replacing sum in (\ref{beta}) by max, we would obtain backward Viterbi recursion (\ref{VAb}) resulting $\max_{u_{1:n}}p(u_{1:n},x_{1:n}).$\\
The forward recursion for marginalization is the following: given the path $x_{1:n}$ define 
\begin{equation}\label{alpha}
\alpha_t(u_t):=p(x_{1:t},u_t),\quad t=1,\ldots,n\end{equation}
and use the following recursion
\begin{equation}\label{alpha-recursion}
\alpha_{t+1}(u_{t+1})=\sum_{u_t}p(x_{t+1},u_{t+1}|x_t,u_t)\alpha_t(u_t).
\end{equation}
Hence $p(x_{1:n})=\sum_{u_n}\alpha_n(u_n).$ Again, replacing the sum by max gives us (\ref{VAf}). Forward and backward recursions are often run simultaneously to obtain the smoothing probability
$$p(u_t|x_{1:n})={\alpha_t(u_t)\beta_t(u_t)\over \sum_{u'_t}\alpha_i(u'_t)\beta_i(u'_t)}.$$
To deal with the numerical underflow, in practice often the scaled versions of the algorithms are used \cite{P04}. However, scaled versions are not required, when one uses Log-Sum-Exp trick \cite{logsumexp}.\\\\
Both methods -- Viterbi and marginalization recursions -- use the distributive property of semirings $(\R_{\geq 0},\max,\cdot)$ and $(\R,+,\cdot)$ respectively. Unfortunately, they can't be combined i.e.
\begin{align*}
\max_{x_{1:n}} \sum_{u_{1:n}} p(x_{1:n},u_{1:n}) &=
\max_{x_{1:n}} \sum_{u_1} p(x_1,u_1) \sum_{u_2} p(x_2,u_2|x_1,u_1) \ldots \sum_{u_n} p(x_n,u_n|x_{n-1},u_{n-1}) \\&\leq
\max_{x_{1}} \sum_{u_1} p(x_1,u_1)  \ldots \max_{x_{n}}\sum_{u_n} p(x_n,u_n|x_{n-1},u_{n-1}),
\end{align*}
which really is a fancy way of writing inequality
\[
\max_x f(x) + g(x) \leq \max_x f(x) + \max_x g(x).
\]
(Note the difference from equality $\max\limits_{x,y} f(x)+g(y)=\max\limits_x f(x)+\max\limits_y g(y)$).

However, we can still use the upper bound obtained by switching the order of summations and maximizations and it's fundamental to the SMS bound introduced in Section \ref{sec:bounds}.

\section{Exact algorithms}\label{sec:exact_algorithms}
In this section, we present a branch-and-bound algorithm and a variety of bounds for solving the maximal marginal problem of a Markov chain.

\subsection{Exhaustive search}
An exhaustive search, also known as a brute-force search, computes the probabilities of all possible sequences. There are $\vert \X \vert^n$ possible sequences, and calculating the probability of each sequence has a time complexity $O(n \vert \U \vert^2)$. Hence, the total time complexity of an exhaustive search is $O(n\,\vert\U\vert^2 \, \vert\X\vert^n)$.

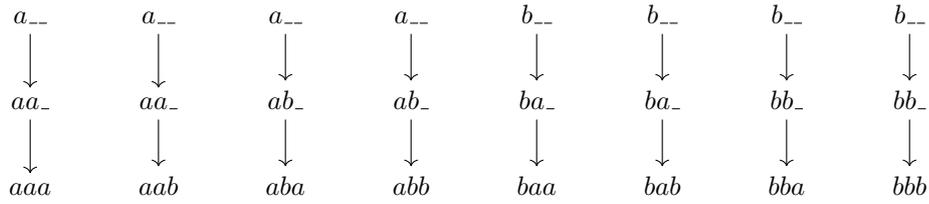
\begin{figure}[h]
\[
\begin{tikzcd}
a\_\_ \arrow[d] & a\_\_ \arrow[d] & a\_\_ \arrow[d] & a\_\_ \arrow[d] & b\_\_ \arrow[d] & b\_\_ \arrow[d] & b\_\_ \arrow[d] & b\_\_ \arrow[d]\\
aa\_ \arrow[d] & aa\_ \arrow[d] & ab\_ \arrow[d] & ab\_ \arrow[d] & ba\_ \arrow[d] & ba\_ \arrow[d] & bb\_ \arrow[d] & bb\_ \arrow[d]\\
aaa & aab & aba & abb & baa & bab & bba & bbb
\end{tikzcd}
\]
\caption{An exhaustive search looking through all possible sequences of length 3 with the alphabet $\X=\{a,b\}$. Informally, this search requires approximately $8 \cdot 3 \vert \U \vert^2 = 24 \vert \U \vert^2 $ operations.}    
\end{figure}

By using divide and conquer, this complexity can be reduced to $O(\vert \U \vert^2 \vert \X \vert^{n+1})$. This improvement is achieved by calculating the probability $p(x_{1:k},u_{1:k})$ for different assignments of $x_k$ by reusing the probability $p(x_{1:k-1},u_{k-1})$ in 
\begin{equation}
\label{eq:markov_rule}
p(x_{1:k},u_k) = \sum_{u_{k-1}} p(x_{1:k-1},u_{k-1}) p(x_k, u_k|x_{k-1}, u_{k-1}).
\end{equation}
\begin{figure}[h]
    \adjustbox{width=\textwidth,center}{
    \begin{tikzcd}
        &     &     &     & \_ \_ \_ \arrow[ld] \arrow[rd] &     &     &     &     \\
        &     &     & a \_ \_ \arrow[ld] \arrow[d] &     & b\_ \_ \arrow[d] \arrow[rd] &     &     &     \\
        &     & aa\_ \arrow[ld] \arrow[lld] & ab\_ \arrow[d] \arrow[ld]    &     & ba\_ \arrow[d] \arrow[rd]   & bb\_ \arrow[rd] \arrow[rrd] &     &     \\
    aaa & aab & aba                         & abb                          &                                & baa                         & bab                         & bba & bbb
    \end{tikzcd}
    }
    \caption{An exhaustive search using divide and conquer to look through all possible sequences of length 3 with the alphabet $\X=\{a,b\}$. Informally, this search requires approximately $14 \cdot \vert \U \vert^2$ operations.}
    \label{fig:exhaustive_search_tree}
\end{figure}
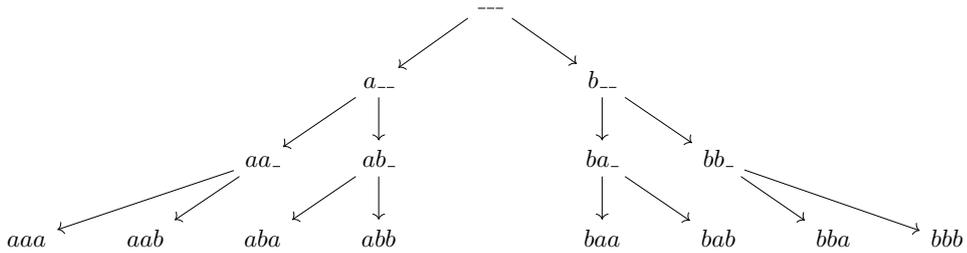

Further improvements can be achieved by eliminating sequences earlier than calculating the probability of the whole sequence -- by using a branch-and-bound algorithm.

\subsection{Branch-and-bound}\label{sec:branch-and-bound}
Branch-and-bound (B\&B) is a method for solving optimization problems by repeatedly dividing the search space into parts (referred to as \textit{branches} due to associated tree structure) and eliminating these parts when possible. Elimination, or \textit{pruning}, is done by calculating bounds for the solution within each branch. If the upper bound of one branch is lower than the lower bound of another, the branch can be pruned from the search space.  

In this case, the search space is the set of all possible sequences $\X^n$. Division into $\vert \X \vert$ branches is accomplished by fixing a state at a specific position in the sequence. For example, consider sequences of length 3 with the alphabet $\X = \{a, b\}$. We can express this set as \_ \_ \_\, akin to the paper-and-pencil game hangman. Fixing the state at the second position results in the two branches: $\_ a \_$ and  $\_ b \_$.
%----------------
\paragraph{Order of fixing positions.}
In this paper, we choose positions in increasing order (i.e. \_ \_ \_ $\rightarrow$ \textasteriskcentered \_ \_ $\rightarrow$ \textasteriskcentered \textasteriskcentered \_ $\rightarrow$ \textasteriskcentered \textasteriskcentered \textasteriskcentered). The corresponding search tree for sequences of length 3 with the alphabet $\X = \{a,b\}$ is illustrated in Figure \ref{fig:exhaustive_search_tree}.

Although decreasing order is equally valid, selecting positions in any arbitrary order (e.g., using a heuristic) presents two main challenges:

First, when states are fixed in non-sequential positions, the effort to calculate the probability of the branch increases. For example, consider a branch $a \_ \_ \_$. To calculate the probability of $a \_ \_ b$, one either uses equation
\[
p(a \_ \_ b) = \sum_{x_{2:3} , u_{1:4}} p(x_1 = a, u_1) p(x_2, u_2 |x_1=a,u_1) p(x_3, u_3 |x_2, u_2) p(x_4=b, u_4 |x_3, u_3)
\]
or some dynamic programming technique. This is in contrast to the sequential ordering, where the segments between consecutive positions always have a size of 0, simplifying calculations to equation \eqref{eq:markov_rule}.

Second, if the position order is not predetermined before the algorithm runs, selecting the next position to fix may require an analysis of all remaining free positions, increasing the algorithm's runtime.

\begin{figure}[h]
    \adjustbox{width=\textwidth,center}{
    \begin{tikzcd}
        &     &     &     &
        \substack{1\\\_ \_ \_\mathstrut\\0.125} \arrow[ld] \arrow[rd] &     &     &     &     
        \\
        &     &     &
        \substack{0.7575\\a \_ \_\mathstrut\\0.1893} \arrow[ld] \arrow[d] &     &
        \substack{0.2425\\b \_ \_\mathstrut\\0.0606} \arrow[d,"/"{anchor=center,sloped}] \arrow[rd,"/"{anchor=center,sloped}] &     &     &     
        \\
        &     &
        \substack{0.3119\\aa\_\mathstrut\\0.1559} \arrow[ld] \arrow[lld] & \substack{0.4456\\ab\_\mathstrut\\0.2228} \arrow[d] \arrow[ld] &     &
        \substack{0.0538\\ba \_\mathstrut\\0.0269} \arrow[d, dashed] \arrow[rd, dashed]   &
        \substack{0.1887\\bb \_\mathstrut\\0.0943} \arrow[rd, dashed] \arrow[rrd, dashed] &     &     
        \\
        \substack{aaa\mathstrut\\0.2887} & \substack{aab\mathstrut\\0.0232} &
        \substack{aba\mathstrut\\0.2128} & \substack{abb\mathstrut\\0.2328} &
             &
        \substack{baa\mathstrut\\0.0316} & \substack{bab\mathstrut\\0.0222} &
        \substack{bba\mathstrut\\0.1461} & \substack{bbb\mathstrut\\0.0426}
    \end{tikzcd}
    }
    \caption{An example of a search with simple bounds. In the case of breadth first search, the algorithm traverses the tree level by level and keeps in memory the best lower bound found so far. The algorithm passes nodes a\_\_, b\_\_, aa\_, ab\_ and then prunes the nodes ba\_, bb\_, because their upper bound is lower than the best lower bound found so far: 0.2228. Then it passes nodes aaa, aab, aba, and abb and chooses the best amongst them: aaa.}
    \label{fig:search_tree}
    % 0.7575 + 0.2425
    % 0.3119 + 0.4456 + 0.0538 + 0.1887
    % 0.2887+0.0232+0.2128+0.2328+0.0316+0.0222+0.1461+0.0426
\end{figure}
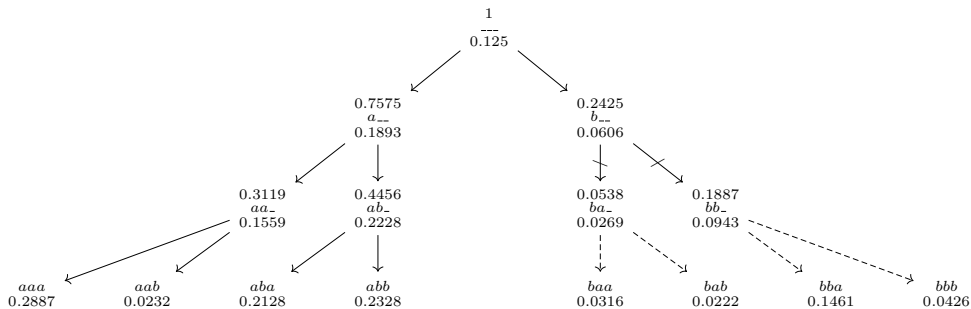

\subsection{Bounds}\label{sec:bounds}
Elimination of the branches of the search tree is done by calculating bounds for the maximal marginal probability in the branch. More precisely, given the beginning of a path $x_{1:k}, (k<n)$, we aim to find upper and lower bounds to the following probability 
$$p^*(x_{1:k})=\max_{x_{k+1}:n}p(x_{1:n}).$$
If an upper bound of $p^*(x_{1;k})$ is smaller than a lower bound of $p^*(x'_{1:k})$ (for some $x'_{1:k}$) the all sequences beginning with 
$x_{1:k}$ are disgarded from the further search. 
\\\\
In the article, the following methods for finding the bounds are used:
\begin{enumerate}
\item \textbf{Simple bounds.} Trivial bounds 
\begin{equation}\label{method:prob}
\frac{p(x_{1:k})}{\vert \X\vert^{n-k}} \leq p^*(x_{1:k})\leq p(x_{1:k}).
\end{equation}
In what follows, bounds (\ref{method:prob}) will be referred to as {\it simple bounds}.

\item \textbf{Power sum bounds.} Find the power sums
$$S_r(x_{1:k})=\sum\limits_{x_{k+1:n}} \big(p(x_{1:n})\big)^r\quad  r\in\N.$$ With $S_r(x_{1:k})$, the  bounds are 
\begin{equation}\label{power}
\sqrt[r]{\frac{S_r(x_{1:k})}{\vert \X\vert^{n-k}}} \leq p^*(x_{1:n}) \leq \sqrt[r]{S_r(x_{1:k})}.
\end{equation}
The inequalities (\ref{power}) hold for trivial reasons: for any tuple of real numbers $a_1\geq \cdots \geq a_N>0$ and any $r \in \N$, it holds $a_1^r<\sum_i a_i^r\leq Na^r_1$. 
Note that $S_1(x_{1:k})=p(x_{1:k})$, so for $r=1$, inequalities (\ref{power}) are the same as (\ref{method:prob}). 
Observe that both lower and upper bound tend to $p^*(x_{1:k})$ as $r$ increases. This observation justifies to use possible big $r$. Unfortunately, the complexity of calculating $S_r(x_{1:k})$ increases with $r$; recursive ways  for computing the power sums are presented in Appendix \ref{sec:power_sum_proof}.
In what follows, the bounds (\ref{power}) will be referred to as {\it power sum bounds} or $r$-PS.

%----------------
\item \textbf{Samuelson type bounds.} The idea is to use $p(x_{1:k})$ and $S_2(x_{1:k})$ simultaneously to bound $p^*(x_{1:k})$. The approach is based on the following extension of Samuelson's inequality \cite{Extensions_of_Samuelsons_Inequality}:
 given $N$ real numbers $a_1 \geq a_2 \geq \cdots \geq a_N$ and power sums $s_1 = a_1 + ... + a_N$ and $s_2 = a_1^2 + \cdots + a_N^2$, 
\begin{equation}\label{eq:samuelson}
\frac{s_1 + \sqrt{\frac{Ns_2-s_1^2}{N-1}}}{N} \leq
a_1 \leq
\frac{s_1 + \sqrt{(N-1)(Ns_2-s_1^2)}}{N}.
\end{equation}
%-------
In our case the numbers $a_i$ are non-negative and then the lower bound above  could be replaced by simpler and typically better bound  $s_2/s_1$. Indeed: for $s_1=1$, it holds
$$s_2=\sum_{i=1}^N a_i^2=a_1^2+(a_2^2+\cdots+a_N^2)\leq a_1^2+(1-a_1)^2\leq a_1.$$
For general $s_1$, now the inequality $s_2/s_1\leq a_1$ trivially follows. The bound $s_2/s_1$ is related to lower bound (\ref{eq:samuelson}): when $s_1^2/s_2$ is an integer then taking $N=s_1^2/s_2$, the lower bound in (\ref{eq:samuelson}) reduces to $s_2/s_1$. The number $M=\left\lceil s_1^2/s_2 \right\rceil$
is the minimal integer such that there exists at least one tuple $a_1\geq \cdots \geq a_M>0$ satisfying $s_1=\sum^M_{i=1} a_i$, $s_2=\sum_{i=1}^M a_i^2$. Plugging $\left\lceil s_1^2/s_2 \right\rceil$ into $N$ in the lower bound of (\ref{eq:samuelson}) gives  $s_2/s_1$.\\
Combining the upper and lower bound, we obtain the following {\it Samuelson type bounds}:
\begin{equation}\label{Sam}
{S_2(x_{1:k})\over p(x_{1:k})}\leq p^*(x_{1:k})\leq {p(x_{1:k}) + 
\sqrt{(|\X|^{n-k}-1)(S_2(x_{1:k})-p(x_{1:k})^2)}\over |\X|^{n-k}}.
\end{equation}

\item \textbf{Swapped max-sum bounds.} Using inequality "$\Max{x}\sum_u \leq \sum_u\Max{x}$" we can switch the order of summations and maximizations to obtain an upper bound which is computationally feasible.  In what follows, the following approach is used: fix a block length $m=1,2,\cdots$. For simplicity, assume for time being that 
$n-k=ml$, where $l\geq 1$ is an integer. We consider the 
probabilities
\begin{align*}
 &p(x_{k+1:k+m},u_{k+m},x_{k+m+1:k+2m},u_{2m}, \ldots, x_{k+(l-1)m:n},u_{k+(l-1)m}|x_{k},u_k)=\\
 &p(x_{k+1:k+m},u_{k+m}|x_k,u_k)p(x_{k+m+1:k+2m},u_{k+2m}|x_{k+m},u_{k+m})\cdots \\
 &\cdots p(x_{k+(l-1)m+1:k+lm}|x_{k+(l-1)m},u_{k+(l-1)m}).
\end{align*}
Observe that summing out $u_{k+jm}$, $j=1,\ldots,l-1$ in the expression above gives
$p(x_{k+1:n}|x_k,u_k)$ and 
$$\sum_{u_k}p(x_{k+1:n}|x_k,u_k)p(u_k, x_{1:k})=p(x_{1:n}).$$
Hence, we can bound  $p^*(x_{1:k})$ above by sum
\begin{align*}
&\sum_{u_k}p(u_k,x_{1:k})\cdot\\
 &\max_{x_{k+1:k+m}} \sum_{u_{k+m}}  p(x_{k+1:k+m},u_{k+m}|u_k,x_k)\cdot\\
 & \max_{x_{k+m+1:k+2m}} \sum_{u_{k+2m}}  p(x_{k+m+1:k+2m},u_{k+2m}|x_{k+m},u_{k+m})\\&\cdots \\
 &
\max_{x_{k+(l-3)m+1:k+(l-2)m}} \sum_{u_{k+(l-2)m}}    p(x_{k+(l-3)m+1:k+(l-2)m},u_{k+(l-2)m}|x_{k+(l-3)m},u_{k+(l-3)m})\cdot \\
& \max_{x_{k+(l-2)m+1:k+(l-1)m}}  \sum_{u_{k+(l-1)m}} p(x_{k+(l-2)m+1:k+(l-1)m},u_{k+(l-1)m}|x_{k+(l-2)m},u_{k+(l-2)m})\cdot \\
 &\max_{x_{k+(l-1)m+1:k+lm}}p(x_{k+(l-1)m+1:k+lm}|x_{k+(l-1)m},u_{k+(l-1)m}). 
\end{align*}
If $(n-k)/m$ is not integer, then we take the first block smaller (i.e. in the second row above $m$ is replaced by $r<m$, where $r+(l-1)m=n-k$). There is no dynamic programming algorithm to maximize over the blocks (of length $m$) in the expression above, but if $m$ is  small, then all probabilities could be calculated and  maximum is easy to find. The intuition suggests that the bigger $m$, the better the upper bound, but as our simulations show, it is not necessarily so. The  dynamic programming algorithm for calculating the approximation is given in Appendix \ref{sec:bounds_by_commutation}. In what follows, the upper bound above will be referred to as {\it $m$-SMS (abbreviation for Swapped Max-Sum bound with blocks of size $m$)}.
%--------------------
%------------------

\item \textbf{$m$-Viterbi approximations.} Let $x'_{k+1:n}$ be an arbitrary path. Clearly $p(x_{1:k},x'_{k+1:n})$ is  a lower bound of $p^*(x_{1:k})$. The bound is good if $x'_{k+1:n}$ is a good approximation of MAP continuation of $x_{1:k}$. There are several computationally cheap ways to obtain an approximation of a Viterbi path, some methods were considered in \cite{KolmekaupaMarkovi}. The results of \cite{KolmekaupaMarkovi} suggests approximating the process $X$ as $m$-th order Markov chain. More precisely, fix $m=1,2,\ldots$ and find the conditional probabilities $p(x_i|x_{i-m:i-1})$, $i=k+1,\ldots,n$. The {\it $m$-Viterbi approximation} calculates a Viterbi path under (usually wrong) assumption that the $X$-process is a $m$-th order Markov chain with the conditional probabilities above. Formally,
\begin{equation}\label{kViterbi}
\check{x}_{k+1:n}=\argmax_{x_{k+1:n}}q(x_{k+1:n}),\end{equation}
where $q$ is the $m$-Markov approximation
\begin{equation}\label{mapp}
q(x_{k+1:n})=\prod_{t=k+1}^np(x_t|x_{t-m:t}).
\end{equation}
In principle, one could use the approximation also when $m<k$, in the present article only the case $m\geq k$ is considered (i.e. the $m$-Viterbi approximation is used only when $k$ is sufficiently big). Observe that with $m=0$, $\check{x}_{k+1:n}$ is just the PMAP-path, i.e. $\check{x}_j=\argmax_{x_j}p(x_j)$, $j=k+1,\ldots,n$.
 The dynamic programming algorithm for calculating the $m$-Viterbi approximation $\check{x}_{k+1:n}$ and the corresponding lower bound $p(x_{1:k},\check{x}_{k+1:n})$ is given in Appendix \ref{sec:k-viterbi_algorithm}.
The intuition suggests that the bigger $m$, the better is the approximation. It is typically so, but according experimental results in \cite{KolmekaupaMarkovi} and by example in Appendix \ref{sec:k-viterbi-not-monotone} it is not always guaranteed. Moreover, it might happen that $m$-Viterbi approximation  has zero probability, for examples see Appendix \ref{sec:k-viterbi_example}.
\\\\
Another meaningful approximation is $\hat{x}_{k+1:n}$, where 
\begin{equation}\label{eq:UX-Viterbi}
(\hat{x}_{k+1:n},\hat{u}_{k+1:n})=\argmax_{(x_{k+1:n},u_{k+1:n})}p(x_{1:k},x_{k+1:n},u_{k+1:n}).
\end{equation}
As explained in Section \ref{sec:dynamic_programming}, $(\hat{x}_{k+1:n},\hat{u}_{k+1:n})$ could be found by Viterbi algorithm. In what follows, the path $\hat{x}_{k+1:n}$ as well as the corresponding lower bound $p(x_{1:k},\hat{x}_{k+1:n})$ will be referred to as {\it UX-Viterbi approximation}.

\end{enumerate}

\section{Experiments}\label{sec:experiments}

The experiments were carried out in the \blackout{High Performance Computing Center of University of Tartu} \cite{UT_Rocket}. The code is available at
%\url{https://gitlab.ut.ee/oskar.soop/article-bb}
.

During experiments we used composite bounding strategies, which means we used selection of bounds from Section \ref{sec:bounds} at once. The overall lower bound for each node $x_{1:k}$ was computed as the maximum of all applied lower bounds, and the upper bound as minimum of all applied lower bounds. By default we always used simple bounds \eqref{method:prob}.

Before conducting further experiments, we evaluated several graph traversal strategies: breadth-first search, depth-first search, and best-first search (for overview on graph traversal strategies read \cite{pearl1984heuristics}). Without incorporating the $m$-Viterbi lower bound, best-first search consistently visited the fewest nodes by a significant margin, with depth-first search performing second best. However, when the $m$-Viterbi lower bound was used, breadth-first search proved to be the most efficient traversal strategy. For larger alphabets, best-first search exhibited performance comparable to breadth-first search. In subsequent experiments, we adopted breadth-first search due to its efficiency and its ability to capture the pruning effect at each layer (see Figure~\ref{fig:layers}).

We summarize the time and space complexities of the algorithms for computing bounds in Table \ref{tab:time_complexities} and \ref{tab:space_complexities}. The preparation time/memory is the time/memory needed to prepare the algorithm for the first node. The time/memory per node is the time/memory needed to calculate the bound for a single node. We provided complexities for two slightly different algorithms for computing the $r$-PS (power sum bounds), as detailed in Appendix~\ref{sec:power_sum_proof}. In the tables we label the algorithm corresponding to equation \eqref{eq:power_sum} as "$r$-PS" and algorithm corresponding to equation \eqref{eq:power_sum_alternative} as "$r$-PS alt".

\begin{table}[htbp]
    \centering
    \caption{The time complexities}
      \begin{tabular}{lcc}
          \hline
            Algorithm & Preparation time & Time per node\\
          \hline
            Simple & $0$ & $O(\vert\U\vert^2)$\\
            $r$-PS & $O(nr\vert\X\vert^2\vert\U\vert^{r+1})$ & $O(r\vert\U\vert^{r+1})$\\
            $r$-PS alt & $\tilde{O}\big(n\vert\X\vert^2 \vert\U\vert^2 \binom{r+\vert\U\vert^2-1}{\vert\U\vert^2-1} r\big)$ & $\tilde{O}\big(\vert\U\vert^2 \binom{r+\vert\U\vert^2-1}{\vert\U\vert^2-1} r\big)$\\
            $m$-SMS & $O(n\vert\X\vert^{m+1}\vert\U\vert^2)$ & $O(m\vert\X\vert^{m-1}\vert\U\vert)$\\
            m-Viterbi & $O(nm\vert\U\times\X\vert^m)$ & $O(1)$\\
          \hline
      \end{tabular}
    \label{tab:time_complexities}
\end{table}

\begin{table}[htbp]
    \centering
    \caption{The space complexities (after computations)}
        \begin{tabular}{lcc}
            \hline
            Algorithm & Preparation memory & Memory per node\\
            \hline
            Simple & $0$ & $O(\vert\U\vert)$\\
            $r$-PS & $O(n\vert\X\vert\vert\U\vert^r)$ & $O(\vert\U\vert^r)$\\
            $r$-PS alt & $O(n\vert\X\vert r^{\vert\U\vert-1})$ & $O(r^{\vert\U\vert - 1})$\\
            $m\text{-SMS}$ & $O(\frac{n}{m} \vert \U\times\X\vert)$ & $O(1)$\\
            $m\text{-Viterbi}$ & $O(n\vert\X\vert^{m-1}\vert\U\vert)$ & $O(1)$\\
            \hline
        \end{tabular}
    \label{tab:space_complexities}
\end{table}

\subsection{Comparing bounds for branch-and-bound algorithm}

We generated 1000 triplet Markov models with $\vert\U \vert = \vert\X\vert = \vert\Y\vert = 2$ and $n=25$. The models were generated by sampling the transition matrices from the Dirichlet distribution with concentration parameter $\alpha=1$. Each model was then used to generate a sequence $y_{1:25}$, which was used to obtain a pairwise Markov model $(U_{1:25},X_{1:25})\vert y_{1:25}$.

For each pairwise Markov model, we executed the breadth-first B\&B algorithm multiple times using different bounding strategies. To evaluate the efficiency of each run, we measured the number of nodes visited, defined as the number of unique paths \( x_{1:k} \) explored. Nodes that were immediately pruned were excluded from this count, ensuring that the minimum possible number of nodes visited equaled the chain length \( n \). Each evaluation used a composite bounding strategy, consisting of a primary bounding method supplemented by the simple bounds.

In the B\&B algorithm, each composite bounding strategy included simple bounds (as defined in equation~\eqref{method:prob}) in addition to a primary bounding method. The overall lower bound was computed as the maximum of all applied lower bounds, and the upper bound as the minimum of all applied upper bounds. We first ran the B\&B algorithm using only the simple bounds, and then augmented it with power sum bounds (\( m \)-PS; see equation~\eqref{power}) for \( m = 2, 5, 10 \), applied to both upper and lower bounds. We also incorporated Swapped Max-Sum bounds (\( m \)-SMS) for \( m = 1, 2, 5, 10 \), applied only as upper bounds.

Subsequently, we evaluated these bounds in combination with the $m$-Viterbi lower bounds. Only the \Viterbi{2} variant was used, as it generally provides a tight approximation to the Viterbi path for short chain lengths. Additionally, we included the Samuelson bound and the 3-PS bound to allow comparison between the Samuelson, 2-PS, and 3-PS bounds. The results are summarized in Table~\ref{tab:results}. For reference, the outcome of an exhaustive search is also shown, though it was not computed. Smallest achievable value in the Table~\ref{tab:results} is $\log_2 25 = 4.6$.

\begin{table}[htbp]
    \centering
    \caption{Number of nodes visited in the B\&B algorithm.}
    \begin{tabular}{lr}
        \toprule
        Method & $\log_2$ of the average \# \\&of nodes visited \\
        \midrule
        exhaustive search & 26.0 \\
        simple & 18.4 \\
        Samuelson & 14.1 \\
        2-PS & 15.4 \\
        3-PS & 13.6 \\
        5-PS & 11.5 \\
        10-PS & 9.1 \\
        1-SMS  & 15.1 \\
        2-SMS  & 14.6 \\
        5-SMS  & 14.3 \\
        10-SMS & 13.4 \\
        simple + 2-Viterbi & 16.5 \\
        2-PS + 2-Viterbi & 12.4 \\
        5-PS + 2-Viterbi & 7.8 \\
        10-PS + 2-Viterbi & 6.2 \\
        1-SMS + 2-Viterbi & 10.0 \\
        2-SMS + 2-Viterbi & 8.5 \\
        5-SMS + 2-Viterbi & 7.0 \\
        10-SMS + 2-Viterbi & 6.1 \\
        \bottomrule
    \end{tabular}
\label{tab:results}
\end{table}

\begin{figure}
    \begin{subfigure}{.5\textwidth}
      \centering
      \includegraphics[width=\linewidth]{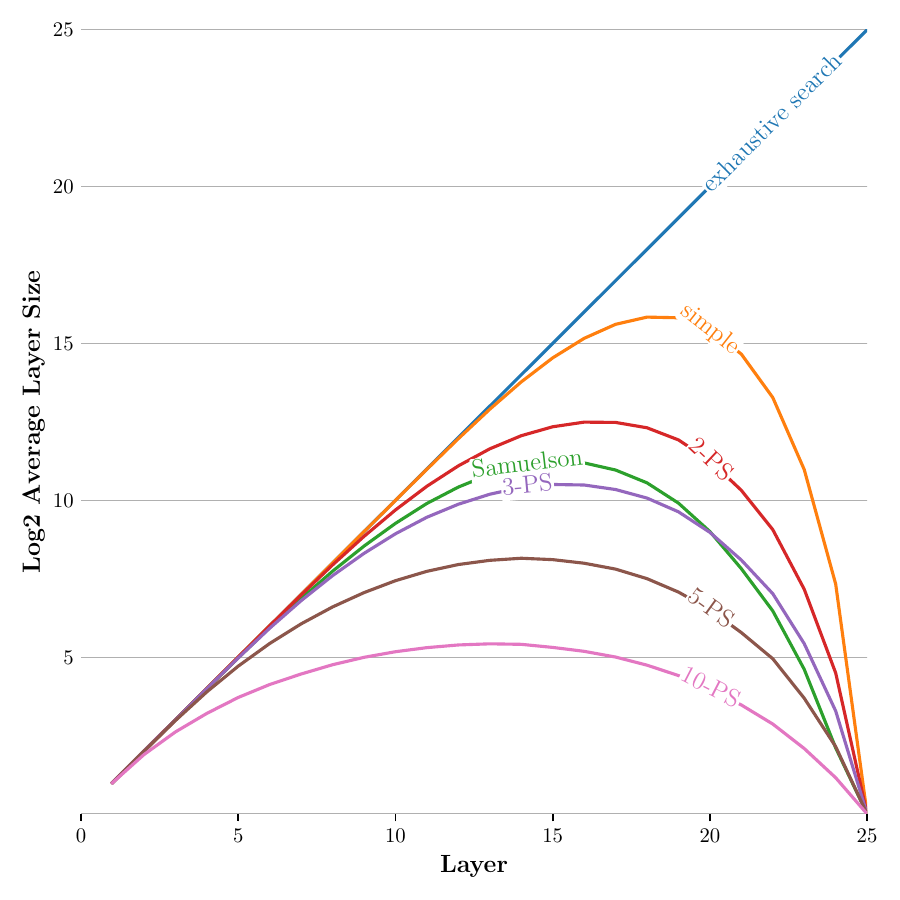}
      \caption{Power sum bounds}
      \label{fig:pfig1}
    \end{subfigure}%
    \begin{subfigure}{.5\textwidth}
      \centering
      \includegraphics[width=\linewidth]{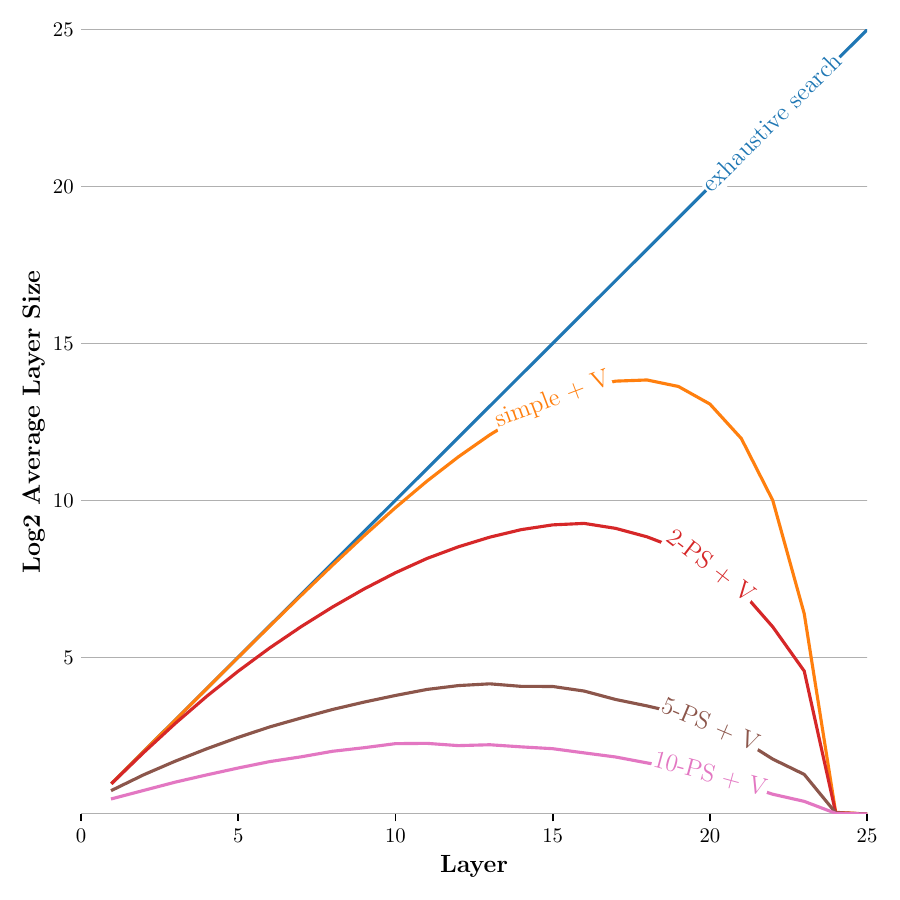}
      \caption{Power sum + 2-Viterbi bounds}
      \label{fig:pfig2}
    \end{subfigure}\\
    \begin{subfigure}{.5\textwidth}
        \centering
        \includegraphics[width=\linewidth]{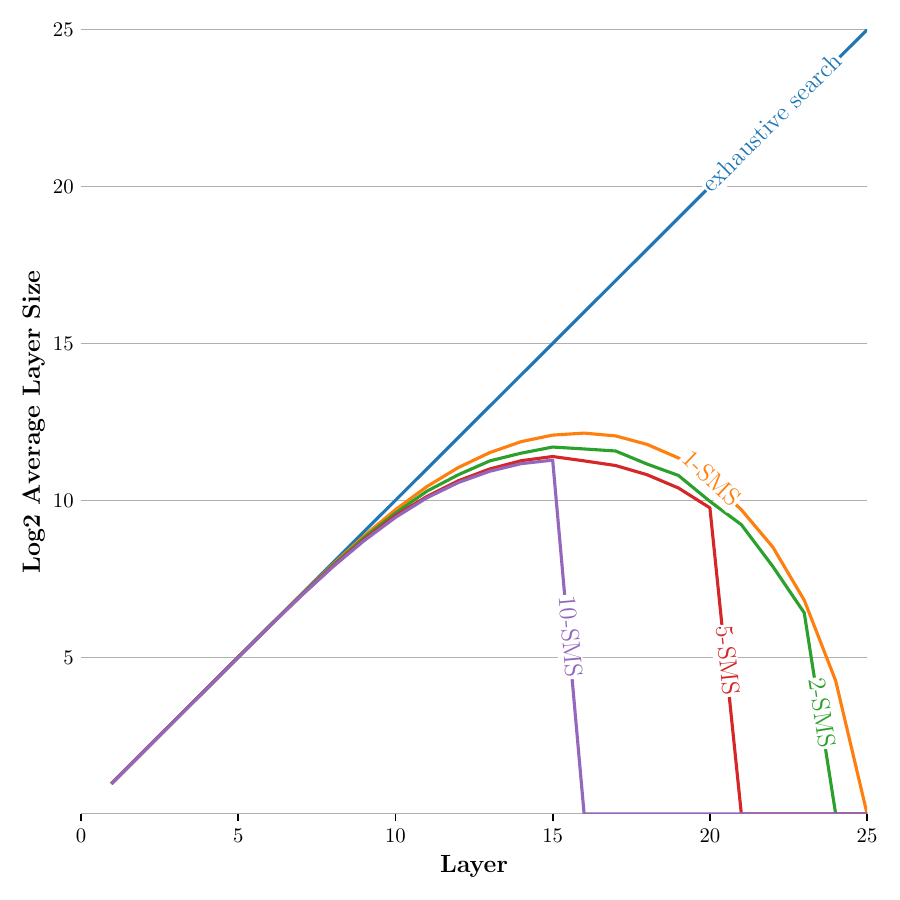}
        \caption{Swapped Max-Sum bounds}
        \label{fig:pfig3}
    \end{subfigure}
    \begin{subfigure}{.5\textwidth}
        \centering
        \includegraphics[width=\linewidth]{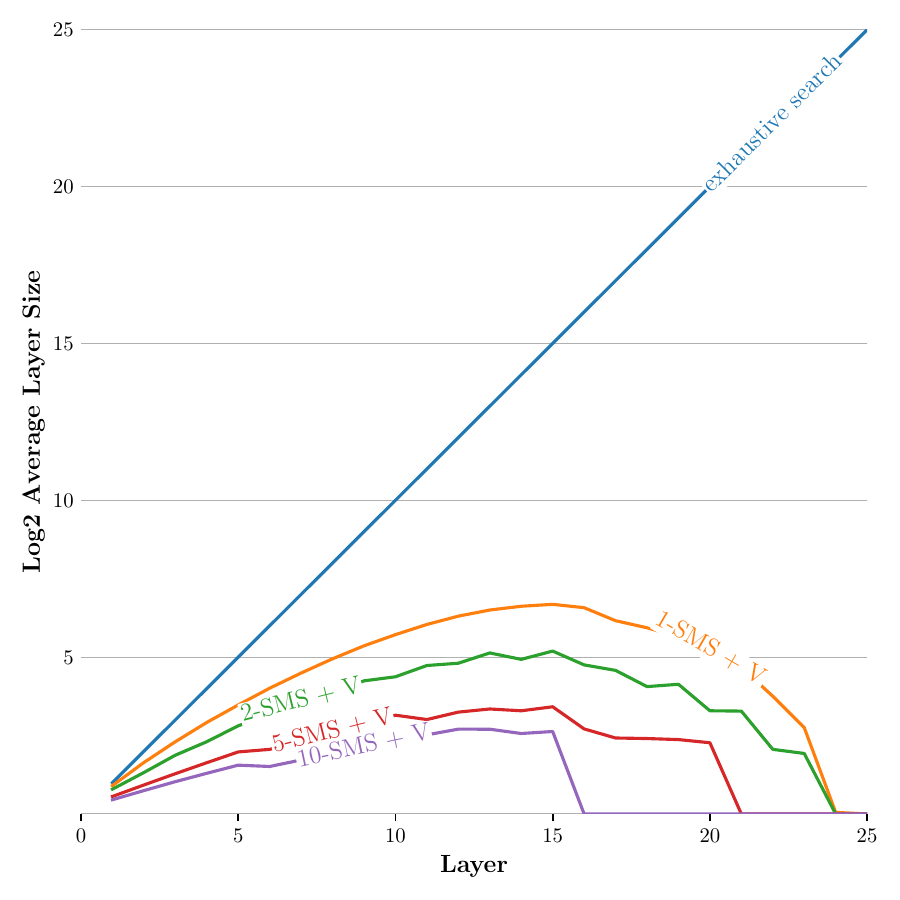}
        \caption{Swapped Max-Sum + 2-Viterbi bounds}
        \label{fig:pfig4}
    \end{subfigure}
    \caption{The number of nodes visited in the B\&B algorithm layer-by-layer.}
    \label{fig:layers}
\end{figure}

It is worth noting that the $m$-SMS method relies on precomputing the final $m$-block of the chain. Therefore, when extrapolating the results to longer chains, it is more appropriate to compare the methods on a layer-by-layer basis, where the $k$-th layer refers to the set of all non-pruned nodes $x_{1:k}$. The corresponding results are presented in Figure~\ref{fig:layers}. The effect of precomputing the final $m$-block can be seen as jumps in Figure \ref{fig:pfig3} and Figure~\ref{fig:pfig4} and to a lesser extent in Figure~\ref{fig:pfig2}.

The results indicate that, among the bounds evaluated, the $m$-Viterbi{} lower bounds should always be included. However, to safeguard against cases where the $m$-Viterbi{} lower bound performs poorly (see Appendix~\ref{sec:k-viterbi_example}), it is advisable to combine it with another lower bound, such as the simple lower bound. No clear winner emerged among the upper bounds. Although the parameters $ m $ in $m$-SMS and $r$ in $r$-PS are not directly comparable, within the class of triplet Markov models considered (i.e., those with $|\X| = |\U| = |\Y| = 2 $), $m$-SMS generally outperformed $r$-PS for small values of $m$ and $r$, and performed similarly or worse for larger values.

\newpage

\subsection{m-Viterbi approximation}

We generated 1000 triplet Markov models with $\vert\U \vert = \vert\X\vert = \vert\Y\vert = 2$ for each $n=100,500,1000$. The models were generated by sampling the transition matrices from the Dirichlet distribution with concentration parameter $\alpha=1$. Each model was then used to generate a sequence $y_{1:25}$, which was used to obtain a pairwise Markov model $(U_{1:25},X_{1:25})\vert y_{1:25}$.

To measure the performance of $m$-Viterbi and UX-Viterbi approximation (and, by extension, the quality of the associated lower bounds), we compared $m$-Viterbi approximation for $m=0,1,2,3,4,5$ and UX-Viterbi against bounds obtained from B\&B algorithm with early stopping. The bounds used in B\&B were composite, consisting of \Viterbi{5}, $5$-SMS and $5$-PS. The algorithm was stopped early when the number of nodes in a layer exceeded $4\cdot 10^5$ or the total number of nodes exceeded $2\cdot 10^6$. These thresholds were chosen to limit the worst-case runtime to approximately one hour and memory usage to around 2 GB.
Since directly computing the difference $p^*-p(\check{x}_{1:n})$ s computationally expensive, we instead assessed accuracy by comparing the log-probability of the $m$-Viterbi approximation $\check{x}_{1:n}$ with the logarithm of the early-stopped bounds. Specifically, we computed $\ln(\text{lower bound})-p(\check{x}_{1:n})$ and $\ln(\text{upper bound})-p(\check{x}_{1:n})$.
The results are summarized in the Table~\ref{tab:results_Viterbi}.

\begin{table}
    \centering
    \caption{Performance of $m$-Viterbi approximation.}
    \begin{tabular}{ccrr}
        \toprule
        Chain's length ($n$) & Parameter ($m$) & \begin{tabular}{@{}c@{}}Average distance \\ from lower bound\end{tabular}  & \begin{tabular}{@{}c@{}}Average distance \\ from upper bound\end{tabular} \\
        \midrule
        100& 0 & 6.6 & 6.7\\
        100& UX & 2.5 & 2.6\\
        100& 1 & 0.9 & 1.0\\
        100& 2 & 0.1 & 0.2\\
        100& 3 & 0.0 & 0.1\\
        100& 4 & 0.0 & 0.1\\
        100& 5 & 0.0 & 0.1\\
        500& 0 & 32.5 & 37.9\\
        500& UX & 11.9 & 17.2\\
        500& 1 & 4.5 & 9.8\\
        500& 2 & 0.6 & 5.9\\
        500& 3 & 0.1 & 5.4\\
        500& 4 & 0.0 & 5.3\\
        500& 5 & 0.0 & 5.3\\
        1000& 0 & 63.5  & 74.4\\
        1000& UX & 24.9 & 35.7\\
        1000& 1 & 9.8 & 20.7\\
        1000& 2 & 1.1 & 12.0\\
        1000& 3 & 0.1 & 11.0\\
        1000& 4 & 0.0 & 10.9\\
        1000& 5 & 0.0 & 10.9\\
        \bottomrule
    \end{tabular}
\label{tab:results_Viterbi}
\end{table}

The results indicate that, within the class of triplet Markov models considered and computational limits, the B\&B algorithm does not significantly improve upon the $m$-Viterbi approximation for $m\geq 2$. The average error $\ln(\text{upper bound}) - \ln(\text{lower bound})$ resulting from early stopping was $0.1, 5.2$ and $10.9$ for $n=100,500$ and $1000$ respectively. 

\section{Conclusion}\label{sec13}

In this article, we have address the computational challenges associated with finding Viterbi paths paths, also known as maximum a posteriori (MAP), in triplet Markov models (TMMs), where standard dynamic programming approaches such as the Viterbi algorithm are no longer applicable due to the loss of the Markov property in marginal processes. Taking advantage of the joint Markov structure of the underlying triplet process, we propose a branch-and-bound framework that incorporates several types of bounds --- simple, power sum (PS), Samuelson-type, Swapped Max-Sum (SMS), and $m$-Viterbi --- to prune the search space efficiently.

Our experimental results demonstrate that while the problem remains NP-hard, the use of tight upper and lower bounds, particularly when combined with m-Viterbi approximations, can drastically reduce the number of candidate paths explored. This approach significantly outperforms exhaustive search in terms of computational complexity and allows for practical computation of Viterbi paths in TMMs of moderate size. Moreover, it enables early stopping during the search process, while still providing rigorous upper and lower bounds on the optimal path probability.

According to our experiments, with simple randomly generated models, no clear winner emerged among the different upper bounds; their effectiveness varies depending on the structure of the problem and available computational resources. We therefore recommend practitioners to experiment with different upper bounding strategies on a case-by-case basis. However, for lower bounds, we recommend using the $m$-Viterbi approximation with $m \geq 2$  as it consistently provided strong performance.

\section*{Declarations}

\blackout{No funds, grants, or other support was received.
The authors have no relevant financial or non-financial interests to disclose.}\\
The code is available at 
%\url{https://gitlab.ut.ee/oskar.soop/article-bb}
.

\begin{appendices}

\section{Power sum formulae}\label{sec:power_sum_proof}

Given the exponent $r\in\N$ and the subsequence $x_{1:k}$ of the sequence $x_{1:n}$, the power sum algorithm computes the sum $S_r(x_{1:k})=\sum_{x_{k+1:n}} p(x_{1:n})^r$. In this section we present two different formulae for the power sum algorithm.

For exponent $r\in\N$ denote by $u^1, u^2, ..., u^r$ independent dummy variables replacing the original sequence $u$ (e.g. $u^2_3$ replaces $u_3$) and denote by $u^{1:r}$ collection of such dummy variables. We use these dummy variables to rewrite expressions like $\big(\sum\limits_u p(x,u)\big)^2$ as $\sum\limits_{u^1,u^2} p(x,u^1)p(x,u^2)$.

The power sum algorithm can be expressed as
\begin{align}
    S_r(x_{1:k}) &=
    \sum_{x_{k+1:n}} p(x_{1:n})^r
    &
    \nonumber \\&=
    \sum_{x_{k+1:n}} \Big(\sum_{u_{k}} p(x_{1:n},u_{k})\Big)^r
    &\text{/marginalization/}
    \nonumber \\&=
    \sum_{x_{k+1:n}} \prod_{i=1}^{r} \sum_{u_{k}^i} p(x_{1:n},u_{k}^i)
    &\text{/introducing dummy variables/}
    \nonumber \\&=
    \sum_{x_{k+1:n}} \prod_{i=1}^{r} \sum_{u_{k}^i} p(x_{1:k},u^i_{k}) p(x_{k+1:n}\vert x_{k},u^i_{k})&\text{/Markov property/}
    \nonumber \\&=
    \sum_{x_{k+1:n}}\sum_{u_k^{1:r}} \prod_{j=1}^r
    p(x_{1:k},u^j_{k}) p(x_{k+1:n}\vert x_{k},u^j_{k})&\text{/distributivity/}
    \nonumber \\&=
    \sum_{u_k^{1:r}}\Big(\prod_{j=1}^rp(x_{1:k},u^j_k)\Big)\Big(\sum_{x_{k+1:n}}\prod_{j=1}^r p(x_{k+1:n}\vert x_{k},u^j_{k})\Big)&\text{/associativity + distributivity/}
    \nonumber \\&=
    \sum_{u^{1:r}_k} \a_k(x_{1:k},u^{1:r}_k) \b_k(x_{k},u^{1:r}_k),
    & \label{eq:power_sum}
\end{align}
where
\begin{equation}\label{eq:PS_alpha_1}
\a_k(x_{1:k},u^{1:r}_k):=\prod_{j=1}^r  p(x_{1:k},u^j_{k})=\prod_{j=1}^r \alpha_k(u_k^j),
\end{equation}
where $\alpha_k(u_k^j)=p(x_{1:k},u^j_k)$ is defined as in (\ref{alpha})
and
\begin{equation}\label{eq:PS_beta_1}
\b_k(x_{k},u^{1:r}_k):=\sum_{x_{k+1:n}}\prod_{j=1}^r p(x_{k+1:n}\vert x_{k},u^j_{k}).
\end{equation}
%---------------
Calculating $\a_k(x_{1:k},u_k^{1:r})$ is straightforward, because $\alpha_k(u_k)$ can be calculated by standard forward-recursion (\ref{alpha-recursion}). It is also possible to use the standard backward-recursion (\ref{beta}) for calculating $p(x_{k+1:n}|x_{k},u_{i})$, but in order to find $\b_k(x_k,u_k^{1:r})$, one has to do it for every possible $x_{k+1:n}$ and then sum over $x_{k+1:n}$, and that is not feasible. We now describe the backward recursion for calculating $\b_k$.\\\\
%-------------
The backward recursion for calculating $\b_t(x_t,u^{1:r}_n)$:
\begin{align*}
    \b_n(x_n, u^{1:r}_n) &= 1
    \\
    \b_i(x_i, u^{1:r}_t) &= \sum_{x_{t+1}, u^{1:r}_{t+1}} p(x_{t+1},u^{1}_{t+1}\vert x_t,u^{1}_t) \cdots p(x_{t+1},u^{r}_{t+1}\vert x_t,u^{r}_t) \beta_{t+1}(x_{t+1},u^{1:r}_{t+1})
    \\
    &= \sum_{x_{t+1}} \sum_{u^{1}_{t+1}} p(x_{t+1},u^{1}_{t+1}\vert x_t,u^{1}_t) \cdots \sum_{u^{r}_{t+1}} p(x_{t+1},u^{r}_{t+1}\vert x_t,u^{r}_t) \beta_{t+1}(x_{t+1},u^{1:r}_{t+1})
\end{align*}
for all $t=1,...,n-1$. The time complexity for calculating $\b_1$ is $O(nr\vert\X\vert^2\vert\U\vert^{r+1})$.
\\\\
Note that for large $r$ and small $\vert\U\vert$ many of the terms $p(x_{t+1},u^1_{t+1}\vert x_t,u^1_t), \dots , p(x_{t+1},u^r_{t+1}\vert x_t,u^r_t)$ are the same. This can be exploited to reduce the number of computations.

Denote $p:=\vert\U\vert$ and $\U=\{a_1,...,a_p\}$ and let 
\[
\Lambda_r := \{(\lambda^1,...,\lambda^p)\in\N^p \,\vert\, \lambda^1+...+\lambda^p=r\}.
\]
%-----
For any $u^{1:r}\in {\cal U}^r$ define $\lambda(u^{1:r})\in \Lambda_r$ as a vector of counts, i.e. $\lambda(u^{1:r})=(\lambda^1,\ldots, \lambda^p)$, where 
$\lambda^l$ denotes how many times $a_l$ appears in $u^{1:r}$ i.e.
\[
\lambda^l(u^{1:r}) = \#\{i\in\{1,...,r\}\vert u^{i}=a_l\}.
\]
Thus we can rewrite equations \eqref{eq:PS_alpha_1} and \eqref{eq:PS_beta_1} as
\begin{align*}
\a_k(x_{1:k},u_k^{1:r})&=\prod_{l=1}^p\big(\alpha_k(a_l)\big)^{\lambda^l_k},
\quad \b_k(x_{k},u^{1:r}_k)=\sum_{x_{k+1:n}}\prod_{l=1}^p\big(p(x_{k+1:n}|x_k,u_k=a_l)\big)^{\lambda^l_k},
\end{align*}
where $(\lambda^1_k,\ldots,\lambda^p_k)=\lambda(u_k^{1:r})$ and $\alpha_k(a_l)=p(x_{1:k},u_k=a_l)$.
\\\\
Replacing $u_k^{1:r}\in|\U|^r$ with $\lambda_k\in \Lambda_r$ we get
$$\a_k(x_{1:k},\lambda_k):= \prod_{l=1}^p\big(\alpha_k(a_l)\big)^{\lambda_k^l},
\quad
\b_k(x_k,\lambda_k):=\sum_{x_{k+1:n}}\prod_{l=1}^p\big(p(x_{k+1:n}|x_k,u_k=a_l)\big)^{\lambda_k^l}$$
with identity
$\a_k(x_k,\lambda_k)=\a_k(x_k,u_k^{1:r})$ and
$\b_k(x_k,\lambda_k)=\b_k(x_k,u_k^{1:r})$ for all vectors $\lambda_k$ and $u_k^{1:r}$ satisfying $\lambda(u_k^{1:r})=\lambda_k$. By counting all the $u_k^{1:r}$ satisfying $\lambda(u_k^{1:r})=\lambda_k$ we obtain
\begin{equation}\label{eq:power_sum_alternative}
S_r(x_{1:k})=\sum_{\lambda_k\in \Lambda_r} \binom{r}{\lambda_k^1,...,\lambda_k^p} \a_k(x_{1:k},\lambda_k)\b_k(x_k,\lambda_k).
\end{equation}

We now describe the backward recursion for calculating $\b_k$. Let $\h{l}{m}_t$ denote how many times pairs $(a_l,a_m)$ appear in $u^{1:r}_{t,t+1}$ i.e. 
\[
\h{l}{m}_t = \#\{i\in\{1,...,r\}\vert u^{i}_{t}=a_l \text{ and } u^{i}_{t+1}=a_m\}
\] and let $h_t$ be a matrix $\big(\h{l}{m}_t\big)_{lm}$. For each $\lambda_t,\lambda_{t+1}\in \Lambda_r$ we define set of such $p\times p$ matrices
\[
H(\lambda_t,\lambda_{t+1}) := \left\{\big(\h{l}{m}_{t}\big)_{lm}\in\N^{p^2} \,\Big\vert\, \h{l}{1}_{t}+...+\h{l}{p}_{t}=\lambda_{t}^l \text{ and } \h{1}{m}_{t}+...+\h{p}{m}_{t}=\lambda_{t+1}^m\right\}.
\]
Fix now $\lambda_t$, $\lambda_{t+1}$ and  $u_t^{1:r}$ such that $\lambda_t=\lambda(u_t^{1:r})$ and let $U(\lambda_{t+1}):=\{u_{t+1}^{1:r}:\lambda(u_{t+1}^{1:r})=\lambda_{t+1}\}$.
Observe that
\begin{align*}
&\sum_{u_{t+1}^{1:r}\in U(\lambda_{t+1})}p(x_{t+1},u^{1}_{t+1}\vert x_t,u^{1}_t) \cdots  p(x_{t+1},u^{r}_{t+1}\vert x_t,u^{r}_t) \b_{t+1}(x_{t+1},u^{1:r}_{t+1})=\\
&\sum_{h_t\in H(\lambda_t,\lambda_{t+1})}\prod_{l=1}^p\binom{\lambda_t^l}{h_t^{l,1},\dots,h_t^{l,p}}\prod_{m=1}^p p(x_{t+1},u_{t+1}=a_m|x_t,u_t=a_l)^{h_t^{l,m}}\b_{t+1}(x_{t+1},\lambda_{t+1}),
\end{align*}
because for fixed $\lambda_t$, $\lambda_{t+1}$ and  $u_t^{1:r}$ we have
\[
\binom{\lambda_{t}^1}{\h{1}{1}_t,\h{1}{2}_t,\dots ,\h{1}{p}_t}
\binom{\lambda_{t}^2}{\h{2}{1}_t,\h{2}{2}_t,\dots ,\h{2}{p}_t}
\cdots
\binom{\lambda_{t}^p}{\h{p}{1}_t,\h{p}{2}_t,\dots ,\h{p}{p}_t}
\]
choices for $u^{1:r}_{t+1}$ resulting in same $h_t$.

Therefore
\begin{align*}
&\sum_{u_{t+1}^{1:r}}p(x_{t+1},u^{1}_{t+1}\vert x_t,u^{1}_t) \cdots  p(x_{t+1},u^{r}_{t+1}\vert x_t,u^{r}_t) \b_{t+1}(x_{t+1},u^{1:r}_{t+1})=\\
&\sum_{\lambda_{t+1}\in \Lambda_r}\sum_{h\in H(\lambda_t,\lambda_{t+1})}\prod_{l=1}^p\binom{\lambda_t^l}{h_t^{l,1},\dots,h_t^{l,p}}\prod_{m=1}^p p(x_{t+1},u_{t+1}=a_m|x_t,u_t=a_l)^{h_t^{l,m}}\b_{t+1}(x_{t+1},\lambda_{t+1})
\end{align*}
and so we obtain the recursion
\begin{align*}
\b_k(x_k,\lambda_k)=
%\sum_{x_{k+1}}\sum_{u_{k+1}^{1:r}}p(x_{k+1},u^{1}_{k+1}\vert x_k,u^{1}_k) \cdots  p(x_{k+1},u^{r}_{k+1}\vert x_k,u^{r}_k) \b_{k+1}(x_{k+1},u^{1:r}_{k+1})\\&=
&\sum_{x_{k+1}}\sum_{\lambda_{k+1}\in \Lambda_r}\sum_{h\in H(\lambda_k,\lambda_{k+1})}\prod_{l=1}^p\binom{\lambda_k^l}{h_k^{l,1},\dots,h_k^{l,p}}
\\&\prod_{m=1}^p p(x_{k+1},u_{k+1}=a_m|x_k,u_k=a_l)^{h_k^{l,m}}\b_{k+1}(x_{k+1},\lambda_{k+1}).
\end{align*}

To calculate the algorithm's time complexity, we can count the number of ways to choose $h_i$. This can be done by utilizing a combinatorial technique called the stars and bars to count the number of $\vert\U\vert\!\times\!\vert\U\vert$ matrices with non-negative integer entries that sum to $r$, resulting in $\binom{r+\vert\U\vert^2-1}{\vert\U\vert^2-1}$ choices for $h_i$. Considering additions and multiplications to be $O(1)$, exponentiation $O(\log r)$ and computation of multinomial be $O(r)$ gives time complexity $O\left(n \vert\X\vert^2 \vert\U\vert^2 \binom{r+\vert\U\vert^2-1}{\vert\U\vert^2-1} r \log{r}\right)$. 
For readability, we may assume that the parameter $|\U|$ is constant, allowing us to replace the binomial coefficient with the bound $O(r^{|\U|^2 - 1})$. Under this assumption, the time complexity simplifies to $O(n \vert\X\vert^2 \vert\U\vert^2 r^{\vert\U\vert^2} \log{r})$, or, by hiding logarithm factor, to $\tilde{O}(n \vert\X\vert^2 \vert\U\vert^2 r^{\vert\U\vert^2})$.

\section{Swapped Max-Sum bounds}\label{sec:bounds_by_commutation}

Using inequality "$\max \sum \leq \sum \max$" we can get upper bounds for the maximal marginal probability by switching the order of maximization and summation. This approach has been used by Park \& Darwiche \cite{Solving_MAP_Exactly_using_Systematic_Search} to solve MAP in Bayesian networks. Considering the Markov chain structure, it is reasonable to generate upper bounds by following rules:
\begin{enumerate}
    \item Start with $\Max{x_1} ... \Max{x_n} \sum_{u_1} ... \sum_{u_n} p(x_{1:n},u_{1:n})=\max_{x_{1:n}}p(x_{1:n})$.
    \item Switch the $\Max{x_s}$ and $\sum_{u_t}$ if they are next to each other and $s\neq t$. There might be multiple choices for this.
    \item Repeat the previous step until satisfied with computational complexity.
\end{enumerate}
Step 2. is motivated by the facts:
\begin{itemize}
\item  We want to keep the temporal ordering: if $s<t$ then $\Max{x_s}$ is to the left of $\Max{x_t}$ and $\sum_{u_s}$ is to the left of $\sum_{u_t}$. This restriction allows to "glue" the cached values from dynamic programming to $p(x_{1:k})$ for every $k$ (see equation \eqref{eq:SMS} below).
\item We want that for every time $t$ maximization $\Max{x_t}$ is to the left of the summation $\sum_{u_t}$, because by swapping their locations such that $\Max{x_t}$ is to the right, computational effort would be similar, but generated bounds would be worse.
\end{itemize}

Every single switch in step 2 increases the upper bound. As an example we can have with $z=(x_{1:3},u_{1:3})$
\begin{align*}
&\max_{x_1} \max_{x_2} \max_{x_3} \sum_{u_1} \sum_{u_2} \sum_{u_3} p(z) \leq
\max_{x_1} \max_{x_2} \sum_{u_1} \max_{x_3} \sum_{u_2} \sum_{u_3} p(z) \\
&\max_{x_1} \max_{x_2} \sum_{u_1} \max_{x_3} \sum_{u_2} \sum_{u_3} p(z) \leq
\begin{matrix}
\max_{x_1} \sum_{u_1} \max_{x_2} \max_{x_3} \sum_{u_2} \sum_{u_3} p(z)\\
\text{or}\\
\max_{x_1} \max_{x_2} \sum_{u_1}  \sum_{u_2} \max_{x_3} \sum_{u_3} p(z).
\end{matrix}
\end{align*}

\begin{figure}
    \includegraphics[width=\textwidth]{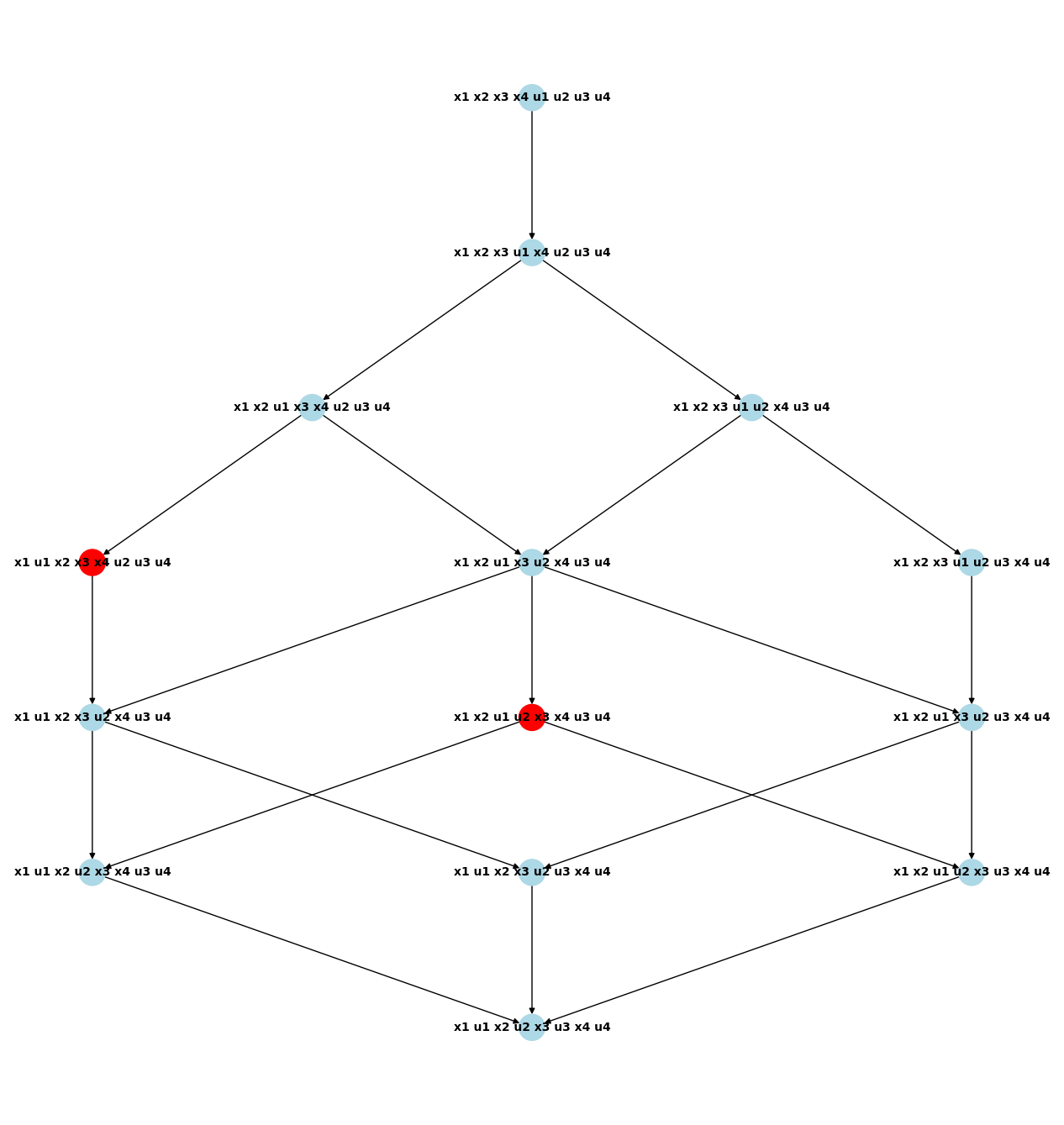}
    \caption{Lattice corresponding to switching the order of maximization and summation in the case of $n=4$. The topmost node is the maximal marginal probability and the bottommost corresponds to 1-SMS bound. Red nodes correspond to 2-SMS and 3-SMS bounds. We can see that these red nodes are not necessarily comparable.}
    \label{fig:lattice_n=4}
\end{figure}

Each ordering of sums and maxes could be viewed as a Dyck word of length $2n$ -- string of $n$ properly opened '(' and closed ')' brackets. For example
$\Max{x_1} \sum_{u_1} \Max{x_2} \Max{x_3} \sum_{u_2} \sum_{u_3}$ would be Dyck word ``()(())''. The corresponding order theoretic lattice is known as Stanley's lattice \cite{StanleyLattice}. The lattice structure for $n=4$ is shown in Figure~\ref{fig:lattice_n=4}. For a fixed pairwise Markov chain, we have a monotone function from the lattice to the real numbers.

We can use these bounds to define $m$-SMS (Swapped Max-Sum bounds with blocks of size $m$) -- the bound obtained by blocks of consecutive maximizations and summations of $m$ variables.

When $z=(x_{1:n},u_{1:n})$, then switching {\it max} and {\it sum} as described above, it is possible step-by-step to reach  to the block-structure as described in Subsection \ref{sec:bounds}: 
\begin{align*}
&\cdots   \max_{x_{n-2m+1}}\cdots \max_{x_{n-m}}     \sum_{u_{n-2m+1}}\cdots \sum_{u_{n-m}}  \max _{_{n-m+1}}\cdots \max_{x_n}\sum_{u_{n-m+1}}\cdots \sum_{u_n} p(z)=\\
&\cdots 
\max_{x_{n-2m+1:n-2m}}\sum_{u_{n-m}}p(x_{n-2m+1:n-m}, u_{n-m}|x_{n-2m},u_{n-2m})
\max_{x_{n-m+1:n}}p(x_{n-m+1:n}|x_{n-m},u_{n-m}).
\end{align*}
Here $m\geq 1$ is the length of block and since it is obtainable with aforementioned switches, we get the upper bound. Since generally the chain's length $n-k$ is not divisible by block size $m$, we take first block of size $r = n-k \: \mathrm{mod}\, m$. We formally leave such smaller block in the beginning of the chain, however it will not be used in the calculations. 

Let $l=\lfloor \frac{n-k}{m} \rfloor$ be number of blocks. Let's define auxiliary function $\delta$ recursively
%----------------------
\begin{align*}
&\delta_{n-m}(x_{n-m},u_{n-m}):=\max_{x_{n-m+1:n}} p(x_{n-m+1:n} \vert x_{n-m}, u_{n-m})\\
&\delta_{n-jm}(x_{n-jm},u_{n-jm}):=\\
& \max_{x_{n-jm+1:n-(j-1)m}}\sum_{u_{k+(j-1)m}}  
p(x_{n-jm+1:n-(j-1)m},u_{n-(j-1)m}|x_{n-jm},u_{n-jm})\cdot \\
&\cdot \delta_{n-(j-1)m}(x_{n-(j-1)n},u_{n-(j-1)m}),\quad j=2,\ldots,l. \\
\end{align*}
When  $r=0$, then $n-lm=k$ and so $\delta_{n-lm}=\delta_k$. When $r>0$, we define also 
$$\delta_{k}(x_k,u_k):=\max_{x_{k+1:k+r}}\sum_{u_{k+r}}p(x_{k+1:k+r},u_{k+r}|x_k,u_k)\delta_{k+r}(x_{k+r},u_{k+r}).$$
%--------------
Observe that $k+r=n-lm$, so that $\delta_{k+r}=\delta_{n-lm}.$ Now clearly
\begin{equation}\label{eq:SMS}
p^*(x_{1:k}) \leq \sum_{u_k}p(x_{1:k},u_k)\delta_k(u_k,x_k).
\end{equation}
The time complexity of computing $\delta$ is $O(n \vert \X\vert^{k+1} \vert \U\vert^2)$ and the time complexity of calculating the upper bound per node is $O(k \vert \U\vert \vert \X\vert^{k-1})$.

\section{The $m$-Viterbi approximation algorithm}\label{sec:k-viterbi_algorithm}
The $m$-Viterbi algorithm approximates the marginal process $X$ with a $m$-th order Markov chain $q$ (recall (\ref{mapp})), where the transition probabilities are the conditional probabilities $p(x_t\vert x_{t-1},...,x_{t-m})$ of the original process $X$.
Calculating these probabilities is possible via forward (\ref{alpha}) or backward (\ref{beta}) recursions. This can be done in $$O\!\left(n\frac{\vert\X\times\U\vert^{m+1}-1}{\vert\X\times\U\vert-1}\right)$$ time.
%-------------------------
Then the Viterbi algorithm is used to find a path  $\check{x}_{k+1:n}$ as in (\ref{kViterbi}). The algorithm is a straightforward generalization of the (backward) Viterbi algorithm (\ref{VAb}): with 
$$\delta_t(x_{t-m:t-1})=\max_{x_{t:n}}q(x_{t:n}|x_{t-m,t-1}),\quad t=k+1,\ldots, n$$
the (backward) recursion is 
\begin{equation}\label{VAm}
\delta_t(x_{t-m:t-1})=\max_{x_{t}}p(x_{t}|x_{t-m,t-1})\delta_{t+1}(x_{t-m+1:t}),\quad t=k+1,\ldots,n
\end{equation}
Here we assume $k\geq m$. Let 
\begin{equation}\label{argmax}
\gamma_t(x_{t-m,t-1})=\arg \max_{x_{t}}p(x_{t}|x_{t-m,t-1})\delta_{t+1}(x_{t-m+1:t}),\quad t=k+1,\ldots,n .\end{equation}
Given sequence $x_{1:k}$, such that $k\geq m$, the $m$-Viterbi approximation $\check{x}_{k+1:n}$ is given by (here $x_{1:k}$ will be denoted as $\check{x}_{1:k}$)
\[
\check{x}_{k+1:n} = (\gamma_{k+1}(\check{x}_{k-m+1:k}), \gamma_{k+2}(\check{x}_{k-m+2:k+1}), ..., \gamma_n(\check{x}_{n-m:n-1})).
\]
%-----------------
Instead of actually finding the sequence $\check{x}_{k+1:n}$, we calculate the probability $p(\check{x}_{k+1:n}\vert x_k, u_k)$. The sequence $\check{x}_{k+1:n}$, and hence the probability, depends only on $x_{k-m+1:k}$ and not the whole $x_{1:k}$.  Let, for any $t=m+1,\ldots,n$
$$\beta_t(x_{t-m,t-1},u_{t-1}):=p(x_{t:n}=\hat{\gamma}(x_{t-m,t-1})_{t:n}|x_{t-1},u_{t-1}),$$
where $\hat{\gamma}(x_{t-m,t-1})_{t:n}\in {\cal X}^{n-t+1}$ is defined  via the functions $\gamma_t$ as in (\ref{argmax}) as follows: with $\hat{\gamma}_j=x_j$ for $j=t-m,\cdots,t-1$, $$\hat{\gamma}_j:=\gamma_j(\hat{\gamma}_{j-m:j-1}),\quad j=t,\ldots,n.$$ Clearly 
\begin{equation}\label{lopp}
p(\check{x}_{k+1:n}|x_{1:k},u_k)=\beta_{k+1}(x_{k-m+1:k},u_k).\end{equation}
The probabilities $\beta_t$ can be calculated via backward recursion as follows: $\beta_{n+1}\equiv 1$ and for $t=k+1,\ldots, n$
\begin{align}\label{abba}
\beta_t(x_{t-m,t-1},u_{t-1})=\sum_{u_t}p(x_{t}= \gamma_{t}(x_{t-m:t-1}), u_t  |x_{t-1},u_{t-1})\beta_{t+1}(x_{t-m+1,t-1},\gamma_{t}(x_{t-m:t-1})).\end{align}
Finally, by (\ref{lopp})
\[
p(x_{1:k},\check{x}_{k+1:n})=
\sum_{u_k} p(x_{1:k},u_k) \beta_{k+1}(x_{k-m+1:k}, u_k).
\]

\section{Examples of bad m-Viterbi approximations}\label{sec:k-viterbi_example}

In this section we will provide examples of cases where $m$-Viterbi approximation behaves much worse than the intuition might suggest. The first example in Subsection \ref{sec:naide0} is a very simple model, where the $m$-Viterbi approximation has 0 probability. The first example assumes that for given $m$  the state space $\U$ is sufficiently large. The second example in Subsection \ref{sec:naide1} is a very simple PMM, with $|\U|=2$, where \Viterbi{1} approximation has positive but exponentially small probability in comparison with the probability of actual Viterbi path.
The third example in Subsection \ref{sec:k-viterbi-not-monotone} exhibits a counter-intuitive model where the probability of $m$-Viterbi approximation decreases when $m$ increases.

 Note that any stochastic process $X$ of length $n$ with finite state space can be modeled as a PMM $(X,U)$ as it suffices to choose $U_i$ to be the history up to this point $X_{1:i-1}$. This means we can choose arbitrary stochastic process as our example.

\subsection{Example of \Viterbi{2} approximation with zero probability}\label{sec:naide0}

Let's consider a stochastic process $X$ of length $n=4$ with state space $\X=\{0,1\}$.
Let the distribution of the process $X$ be
\[
p(0,1,1,1) = p(1,0,1,1) = p(1,1,0,1) = p(1,1,1,0) = \frac{1}{4}.
\]
Hence
\[
p(x_i=1)=\frac{3}{4}, \quad p(x_i=0)=\frac{1}{4},
\]
\[
p(x_i=1\vert x_{i-1}= 1) = \frac{2}{3}, \quad p(x_i=0\vert x_{i-1}=1) = \frac{1}{3}, \quad p(x_i=1\vert x_{i-1}=0) = 1
\]
and
\begin{align*}
&p(x_i=1\vert x_{i-2}=1,x_{i-1}=1) = \frac{1}{2},
&p(x_i=0\vert x_{i-2}=1,x_{i-1}=1) = \frac{1}{2},\\
&p(x_i=1\vert x_{i-2}=1,x_{i-1}=0) = 1, 
&p(x_i=1\vert x_{i-2}=0,x_{i-1}=1) = 1.
\end{align*}
We now calculate the Viterbi (MAP) paths of $q(x_{1:4})$, where $q$ is $m$-order Markov approximation of true measure $p$. We consider the following values $m=0,1,2$. 

For $m=0$, the measure $p$ is approximated by product measure (independence) and so the
\Viterbi{0} approximation (PMAP-path) is
\[
\argmax_{x_{1:4}} \prod_{i=1}^4 p(x_i) = (1,1,1,1),
\]
For $m=1$, the measure $p$ is approximated by (1-order) Markov chain and so the \Viterbi{1} approximation is
\[
\argmax_{x_{1:4}} p(x_1)\prod_{i=2}^4 p(x_i\vert x_{i-1}) = (1,1,1,1),
\]
For $m=2$, the measure $p$ is approximated by second-order Markov chain and \Viterbi{2} approximation is
\[
\argmax_{x_{1:4}} p(x_1,x_2)\prod_{i=3}^4 p(x_i\vert x_{i-1}, x_{i-2}) = \{(1,0,1,1), (1,1,0,1)\}.
\]

% \begin{align*}
% \hat{p}(0,1,1,0) = \frac{1}{4} 1 \frac{1}{2} = \frac{1}{8}\\
% \hat{p}(0,1,1,1) = \frac{1}{4} 1 \frac{1}{2} = \frac{1}{8}\\
% \hat{p}(1,0,1,1) = \frac{1}{4} 1 1 = \frac{1}{4}\\
% \hat{p}(1,1,0,1) = \frac{1}{2} \frac{1}{2} 1 = \frac{1}{4}\\
% \hat{p}(1,1,1,0) = \frac{1}{2} \frac{1}{2} \frac{1}{2} = \frac{1}{8}\\
% \hat{p}(1,1,1,1) = \frac{1}{2} \frac{1}{2} \frac{1}{2} = \frac{1}{8}
% \end{align*}

As we can see, the \Viterbi{0} and \Viterbi{1} approximations have probability zero and \Viterbi{2} approximation provides an exact solution.

The example process $X$ is uniquely defined by the location of the zero and therefore the process can be modeled as PMM if $\vert\U\vert \geq 4$. The example extends to any $n > 4$ by appending $n-4$ ones to the sequence.\\\\

This example generalizes to a stochastic process of arbitrary length $n$, where sequences are equiprobable and consist of of $n-1$ ones and one zero. Such process can be modeled as a PMM with $|\U|\geq n$. Rudimentary analysis can show that for $n>8$ if $2m<n$, then $m$-Viterbi approximation has probability zero.

\subsection{An example with $\vert\X\vert=\vert\U\vert=2$} \label{sec:naide1}
The following example shows that the \Viterbi{1} approximation can be with exponentially low probability (in $n$) even when $\vert\X\vert=\vert\U\vert=2$. We are going to construct a PMM  with typical realization  like this
\[
\begin{pmatrix}
    u_{1:n}\\
    x_{1:n}
\end{pmatrix}
=
\begin{pmatrix}
    1 & 1 & 1 & 0 & 0 & ... & 0\\
    1 & 1 & 1 & 0 & 1 & ... & 1
\end{pmatrix},
\]
where $U$ is a stream of ones which flips to zeros at some random point in time and $X$ shows where the flip happens. More precisely, let $(X,U)$ be a PMM with initial probabilities being  $p(x_1 =1, u_1 = 1) = p,$  $p(x_1 = 0, u_1 = 0) = 1-p$ (thus $X_1=U_1$) and the transition probabilities being 
\begin{align*}
&p(x_i,u_i|x_{i-1},u_{i-1})= p(u_i|u_{i-1})p(x_i|u_i,u_{i-1}),\quad   \text{where}\\ 
&p(x_i = 0 \vert u_i=0, u_{i-1}=1) = 1, \quad p(x_i = 1 \vert u_i= u_{i-1}) = 1.
\end{align*}
This type of PMM-s are known as {\it Markov switching models} \cite{hyb} and it is easy to see that now $U$ is a Markov chain with
 $p(u_1 = 1) = p$ and transitions
\[
p(u_i=1 \vert u_{i-1}=1) = p, \quad p(u_i=0 \vert u_{i-1}=1) = 1-p, \quad p(u_i=0 \vert u_{i-1}=0) = 1
\]
(a left-right Markov chain with 0 being absorbing state). Then the probability of the sequence $x_{1:n}$ is
\[
p(x_{1:n}) = \begin{cases}
    p^n, & \text{if $x_{1:n}$ has no zeros}\\
    p^{i-1}(1-p), & \text{ if $x_i$ is the only zero}\\
    0, & \text{otherwise}.
    \end{cases}
\]
For sufficiently large $n$, the most likely sequence is 0,1,1,...,1 with probability $1-p$.

To find \Viterbi{1} approximations we find the conditional probabilities
\begin{align*}
&p(x_i = 0 \vert x_{i-1} = 1) = {p(x_{i-1}=1,x_i=0)\over p(x_{i-1}=1)}= {p^{i-1}(1-p)\over 1-p^{i-2}(1-p)}, \\
&p(x_i = 1 \vert x_{i-1} = 1) = 1 -{p^{i-1}(1-p)\over 1-p^{i-2}(1-p)}={1-p^{i-2}(1-p^2)\over 1-p^{i-2}(1-p)},\\
&p(x_i = 1 \vert x_{i-1} = 0) = 1
\end{align*}
and note that \Viterbi{1} approximation can't have two zeros in sequence $x_{i-1}=x_{i}=0$.

Given arbitrary sequence of ones and zeros, where there are no two zeros next to each other, by switching 1 to 0 at position $i$, the approximant objective value changes multiplicatively by
\[
\frac{
    p(x_i = 0 \vert x_{i-1} = 1)p(x_{i+1} = 1 \vert x_{i} = 0)
    }{
    p(x_i = 1 \vert x_{i-1} = 1)p(x_{i+1} = 1 \vert x_{i} = 1)
    }
=
\frac{p^{i-1}(1-p)}{\big(1-p^{i-2}(1-p^2)\big)} {(1-p^{i-1}(1-p))\over (1-p^{i-1}(1-p^2))}, \quad i\geq 2.
\]
For $i\geq 2$ this ratio is less than 1, when $p>\gamma_1$, where $\gamma_1\approx0.550$.
For $i=1$
$${p(x_1=0)p(x_2=1|x_1=0)\over p(x_1=1)p(x_2=1|x_1=1)}={(1-p)\over p^2}, $$
because
$p(x_2=1|x_1=1)=p.$ The ratio is less than 1, when  $p>\gamma_2$, where $\gamma_2\approx 0.618$. This means that for sufficiently big $p$, the \Viterbi{1} approximation is $1,1,...,1$, which has probability $p^n$. So, when $p>\gamma_2$, the probability of \Viterbi{1} approximation can be  arbitrary small ($p^n$), while the probability of the true Viterbi path remains constant $(1-p)$.

\subsection{An example when 0-Viterbi is better than 1-Viterbi}\label{sec:k-viterbi-not-monotone}
The following example shows that $m<m'$ does not imply that $m'$-Viterbi is better than $m$-Viterbi approximation.

Consider set of possible outcomes 111, 100, 101, 001, 011, 010 with odds $1+\varepsilon:1:1:1:1:1$.
Then
\[
p(x_1=1)=\frac{3+\varepsilon}{6+\varepsilon}, \quad
p(x_2=1)=\frac{3+\varepsilon}{6+\varepsilon}, \quad
p(x_3=1)=\frac{4+\varepsilon}{6+\varepsilon}
\]
and
\begin{align*}
&p(x_2=1 | x_1=1)=\frac{1+\varepsilon}{3+\varepsilon},  &p(x_2=0 | x_1=1)=\frac{2}{3+\varepsilon}\\
&p(x_2=1 | x_1=0)=\frac{2}{3}, &p(x_2=0 | x_1=0) = \frac{1}{3}
\\
&p(x_3=1 | x_2=1)=\frac{2+\varepsilon}{3+\varepsilon},  &p(x_3=0 | x_2=1)=\frac{1}{3+\varepsilon}\\
&p(x_3=1 | x_2=0)=\frac{2}{3}, &p(x_3=0 | x_2=0) = \frac{1}{3}.\\
\end{align*}
Obviously 0-Viterbi results in the most likely sequence 111.
The Markov approximation is
\begin{align*}
q(100) &= \frac{3 + \varepsilon}{6 + \varepsilon} \cdot \frac{2}{3 + \varepsilon} \cdot \frac{1}{3}
= \frac{2}{3(6 + \varepsilon)} 
&&
q(001) = \frac{3}{6 + \varepsilon} \cdot \frac{1}{3} \cdot \frac{2}{3}
= \frac{2}{3(6 + \varepsilon)} \\[1.5ex]
q(111) &= \frac{3 + \varepsilon}{6 + \varepsilon} \cdot \frac{1 + \varepsilon}{3 + \varepsilon} \cdot \frac{2 + \varepsilon}{3 + \varepsilon}
= \frac{(1 + \varepsilon)(2 + \varepsilon)}{(6 + \varepsilon)(3 + \varepsilon)}
&&
q(010) = \frac{3}{6 + \varepsilon} \cdot \frac{2}{3} \cdot \frac{1}{3 + \varepsilon}
= \frac{2}{(6 + \varepsilon)(3 + \varepsilon)} \\[1.5ex]
q(101) &= \frac{3 + \varepsilon}{6 + \varepsilon} \cdot \frac{2}{3 + \varepsilon} \cdot \frac{2}{3}
= \frac{4}{3(6 + \varepsilon)}
&&
q(011) = \frac{3}{6 + \varepsilon} \cdot \frac{2}{3} \cdot \frac{2 + \varepsilon}{3 + \varepsilon}
= \frac{2(2 + \varepsilon)}{(6 + \varepsilon)(3 + \varepsilon)} \\[1.5ex]
q(110) &= \frac{3+\varepsilon}{6+\varepsilon} \cdot \frac{1+\varepsilon}{3+\varepsilon} \cdot \frac{1}{3+\varepsilon} = \frac{1+\varepsilon}{(3+\varepsilon)(6+\varepsilon)}
&&
q(000) = \frac{3}{6+\varepsilon} \cdot \frac{1}{3} \cdot \frac{1}{3} = \frac{1}{3(6+\varepsilon)}.
\end{align*}
For small $\varepsilon$ we have that $q(101)\approx q(011) \approx \frac{4}{18}$, but $q(111)\approx\frac{2}{18}$.
Hence the \Viterbi{1} approximation has probability $1/(6+\varepsilon)$, but \Viterbi{0} approximation is the actual Viterbi (MAP) path with higher probability $(1+\varepsilon)/(6+\varepsilon)$.

\end{appendices}


\bibliography{reference}% common bib file

\end{document}